\documentclass[epj,nopacs]{svjour}
\usepackage{graphics}
\usepackage{array}
\usepackage{multirow}
\usepackage{amssymb}
\usepackage{amsmath}
\usepackage{array}
\usepackage{tablefootnote}
\usepackage{tikz}
\usepackage{pgfplots}
\usepackage{xcolor}
\usepackage{subcaption}
\usepackage{url}
\usetikzlibrary{arrows}
\definecolor{plotpink}{rgb}{1,0,1}
\definecolor{plotpurple}{rgb}{0.5,0,0.5}
\usepackage{xcolor}

\begin{document}
\title{Constraints and Entropy in a Model of Network Evolution}

\author{Philip Tee \inst{1,2} \and Ian Wakeman \inst{2} \and George Parisis \inst{2} \and Jonathan Dawes \inst{3} \and Istv\'an Z. Kiss \inst{4}}

\institute{%
 Moogsoft Inc, \\ 1265 Battery Street, San Francisco, California 94111 USA
 \email{phil@moogsoft.com}
 \and
 School of Engineering and Informatics, \\ University of Sussex, Brighton BN1 9RH UK
 \email{{ianw,g.parisis,p.tee}@sussex.ac.uk}
 \and
 Centre for Networks and Collective Behaviour, and \\
Department of Mathematical Sciences, \\
University of Bath, Bath BA2 7AY, UK
\email{j.h.p.dawes@bath.ac.uk}
\and
Department of Mathematics, School of Mathematical and Physical Sciences, \\ University of Sussex, Brighton BN1 9QH, UK
\email{i.z.kiss@sussex.ac.uk}
}%

\abstract{
Barab\'{a}si-Albert's `Scale Free' model is the starting point for much of the accepted theory of the evolution of real world communication networks. Careful comparison of the theory with a wide range of real world networks, however, indicates that the model is in some cases, only a rough approximation to the dynamical evolution of real networks. In particular, the exponent $\gamma$ of the power law distribution of degree is predicted by the model to be exactly 3, whereas in a number of real world networks it has values between 1.2 and 2.9. In addition, the degree distributions of real networks exhibit cut offs at high node degree, which indicates the existence of maximal node degrees for these networks.
In this paper we propose a simple extension to the `Scale Free' model, which offers better agreement with the experimental data. This improvement is satisfying, but the model still does not explain \emph{why} the attachment probabilities should favor high degree nodes, or indeed how constraints arrive in non-physical networks. Using recent advances in the analysis of the entropy of graphs at the node level
we  propose a first principles derivation for the `Scale Free' and `constraints' model from thermodynamic principles, and demonstrate that both preferential attachment and constraints could arise as a natural consequence of the second law of thermodynamics.
} 
\maketitle
%


\section{Introduction and Background}
\subsection{Overview}

The \emph{`Scale Free'} model of Barab{\'a}si-Albert \cite{Albert2002} is widely accepted as the definitive model of how real world networks evolve. This and other dynamic network models consider real world networks as graphs $G(V,E)$, where $V(t)$ is the set of vertices and $E(t)$ the set of edges.
Its success at overcoming the difficulties of applying the  Erd\H{o}s-R{\'e}nyi (ER) random graph model (for a detailed description see \cite{Bollobas2001}) to real world networks is well understood. 
In particular the model naturally results in a power law degree distribution, as opposed to the random graph model, which has a binomial distribution of node degree, which in the continuum limit of a very large network is approximately Poisson, with well defined higher statistical moments that establish the `scale' of the graph. This is in stark contrast to the scale free model which does not have well defined moments above the mean. 
The model described by Barab{\'a}si-Albert in \cite{Albert1999} and \cite{Albert2002} builds upon, and provides an explanation for, the notion of the small world network, first introduced by Watts and Strogatz \cite{Watts1998} and has been used to analyze a wide variety of real world graphs.

On close examination, the scale free model has a number of theoretical challenges, and, it is well understood that the behavior of real world networks has deeper complexity than a single constant power law degree distribution. Of course balanced against the success of the model in generating networks that share the \emph{small world} property and scale free degree distributions, these challenges can be viewed as opportunities for refinement of the fundamental approach. In this work we focus on extensions to the model which provide improvements in the following three areas:

\begin{itemize}
	 \item \emph{Absence of Constraints:} There is an assumption that a graph can continue to evolve indefinitely, unconstrained by any system wide or external resources. For most real world networks this is not the case. For example in communication networks every node in the network has a natural maximum connectivity. In the scale free model there is no such upper limit to node degree.
    \item \emph{Fit to Real World Data:} The standard scale free model produces a degree distribution that follows a power law with exponent $\gamma=3$. It is well understood that this is not an exact fit to real world data, which we highlight in Section \ref{sec:analysis}. Many extensions exist that produce a better fit, some of which we survey later. It is clear that the degree distributions of real networks have more complex behavior than a simple fixed exponent power law.
    \item \emph{Absence of a Physical Model:} The notion of scale freedom derives directly from the hypothesis of preferential attachment, that is in a dynamically evolving graph new nodes will more likely attach to nodes of higher degree. Whilst the scale free model provides a theoretical framework that points to high node degree making a node more likely to attract new connections, there is no fundamental explanation of \emph{why} that should be so, and what physical processes may be at work that could produce that effect. It would be desirable if this could be explained using a first principles argument involving well understood mechanisms. This would further strengthen the fundamental premise of the scale free model.
\end{itemize}

In this paper we will attempt to address these challenges. We do so by proposing a simple extension to the standard scale free model, which introduces a hard cut off in the degree of a node, motivated by considerations from communications network design. This model has some attractive features, amongst which is a more accurate prediction of the power law exponent. Although extensions to the preferential attachment approach (most notably \cite{Dorogovtsev2000}, \cite{Bianconi2001} and \cite{Krapivsky2000}), can result in values of the power law exponent less than $3$, we believe our model achieves this through a simple and natural extension to the traditional preferential attachment paradigm. Furthermore, as a consequence of introducing the constraint, we identify that the attachment probability introduces \emph{superlinear} polynomial terms in node degree. This additional structure to the attachment probability is responsible for a richer scaling regime in node degree evolution. This structure allows us to compare in Section \ref{sec:dynamics} both the constraints and scale free model to a novel model of evolution that argues from a stochastic perspective based upon recent developments in the structural entropy of a graph. By developing the outline of an entropic model we illustrate how both the standard scale free and our constrained model could be viewed as approximations to a more fundamental, statistical thermodynamic model of network growth.

In this section we will begin with a brief overview of the continuum analysis used in \cite{Albert2002} to derive the principle results of scale free models, and at a very high level subsequent attempts to build upon and extend the model. We will make use of the same continuum approximation in our analysis.

We show in section \ref{sec:constraints} how the introduction of a simple environmental constraint into the scale free model can significantly improve its predictive power, and compare our \emph{constrained} model to a range of more contemporary network data in section \ref{sec:analysis}. As part of the verification of our \emph{constrained} model, we also present results of simulations of network growth using our modified attachment probability defined in Section \ref{sec:constraints}.
An attractive feature of our extended model is that it reproduces the scale free model when we allow our constraint to tend to infinity. We are able to significantly outperform the ability of the scale free model to predict the exponent $\gamma$ of the power law distribution across a wide range of real world data (results are summarized in Table \ref{tab:gammatab}). 
In particular for ten of the twenty three data sets analyzed (marked in Table \ref{tab:gammatab} in bold) we are able to predict $\gamma$ to within 10\%, whereas the scale free model  overestimates the value of $\gamma$ by an average of 35\% and in only four cases does it predict within the range 10-20\%.
Our constrained model therefore performs better than the standard scale free model on the first two issues identified above, but not on the third.

In Section \ref{sec:dynamics} we propose a novel statistical thermodynamical (i.e. entropic) model of network growth. This addresses the third objective. Recent work on the behavior of communications networks by Tee \emph{et al} \cite{Tee2016b,Tee2017} introduced a measure of the structural entropy of a node, derived from its degree and clustering coefficient. We show how this can lead to a direct derivation of scale free and constraint models, potentially explaining why scale freedom arises and why our constrained model is a better fit for networks as they grow and encounter connectivity limitations. We present in the same section some early results from numerical simulations of the entropic model, which show many of the features of the real world data we analyzed in Section \ref{sec:analysis}.
\subsection{The Scale Free Model}
The Scale Free Model of Barab{\'a}si, Albert and Jeong \cite{Albert1999}, \cite{Albert2002} is based on two simple and fundamental assumptions:

\begin{itemize}
	\item \emph{Growth:} Starting with $m_{0}$ nodes and $e_{0}$ edges, we add a new node at each unit time step. When this node is added to the network, it connects to $m \ll m_{0}$ other nodes. This process continues indefinitely, such that after $t$ unit time steps, there are $m_{0}+t$ nodes, and $e_{0} + mt$ edges. Eventually the constants in these expressions can be dropped as they are insignificant compared to $t$.
	\item \emph{Preferential Attachment:} The node attaches to other nodes with a probability determined by the degree of the target node, such that more highly connected nodes are \emph{preferred} over lower degree nodes.
\end{itemize}

Using a mean field theory approach the analysis explains both the power law scaling of real world networks \cite{Faloutsos1999}, and the simultaneous resilience and vulnerability of networks to random and targeted attacks, respectively \cite{Albert2000}. The approach taken in \cite{Albert1999} begins by proposing the probability of a \emph{randomly chosen node $i$}, capturing a connection to a new node, as solely dependent upon its degree $k_{i}$ as:

\begin{equation}\label{eqn:pref_pi}
	\Pi_{i}=\frac{k_{i}}{ \sum_{j} k_{j} } = \frac{k_{i}}{2mt}  \mbox{ ,}
\end{equation}

In the strictest sense the approximation $\sum_{j} k_{j} = 2mt$ should include the original nodes $m_{0}$ and their degrees, however for large values of $t$ this can be effectively ignored, without loss of generality, as $2e_{0} \ll 2mt$. By taking the continuous approximation, this naturally leads to the following ordinary differential equation for the time evolution of node $i$'s degree $k_{i}(t)$:

\begin{equation}\label{eqn:pref_attachment}
	\frac{d{k_{i}(t)}}{dt}=m\Pi_{i}=\frac{k_{i}(t)}{2t} \mbox{ .}
\end{equation}

Equation (\ref{eqn:pref_attachment}), can be solved subject to an initial condition that at time $t_{i}$, when node $i$ is added, its degree $k_{i}=m$ to yield:

\begin{equation}
	k_{i}(t)=m \Bigg(  \frac{t}{t_{i}}\Bigg)^{1/2} \mbox{ .} \label{eqn:ba}
\end{equation}

In order to derive the degree distribution begin by assuming that $t$ is fixed. At this stage the probability that $k_{i}(t)$ is smaller than a given degree $k$ is:

\begin{equation*}
	P( k_{i}(t) < k  ) = P \Bigg( t_{i} > \frac{m^{2}t}{k^{2}} \Bigg)  = 1 - P \Bigg( t_{i} \leq \frac{m^{2}t}{k^{2}} \Bigg) \mbox{ .}
\end{equation*}

Developing the mean field approach we note that the $i$th node was chosen at random, so its time of introduction into the network $t_{i}$ is a random variable. Given that nodes are added at each time step, the range of possible values for $t_{i}$ are $1,2,\dots,(m_{0}+t)$, and each value can occur with probability $\frac{1}{(m_{0}+t)}$. We can conclude that the random variable $t_{i}$ is uniformly distributed and can write the probability of choosing a node $i$ with a $t_{i}$ smaller than $\frac{m^{2}t}{k^{2}}$ as:

\begin{equation*}
	 P \Bigg( t_{i} \leq \frac{m^{2}t}{k^{2}} \Bigg) = \frac{1}{(m_{0}+t)} \times \frac{m^{2}t}{k^{2}} \mbox{ .}
\end{equation*}

We can now state that the probability of a node having degree $k < k_{i}$ as:

\begin{equation*}
	P( k_{i}(t) < k  ) = 1 - \frac{m^{2}t}{k^{2}(m_{0}+t)} = \int_{m}^{k}P(x) dx \mbox{ ,}
\end{equation*}

implying that $P(k) = \frac{ \partial P( k_{i}(t) < k  ) }{ \partial k }$, yielding the principal result of the Barab{\'a}si-Albert Scale Free model:

\begin{equation}
	P(k) = \frac{2m^{2}t}{m_{0} + t } \frac {1}{k^{3}}\mbox{ .}
\end{equation}

This predicts that on a log/log scale the slope of the degree distribution $\gamma$ is identically $3$. The result has been compared against many real world networks, and indeed the power law behavior has been seen in many examples and is one of the triumphs of the scale free model. The model, however, generally overestimates the value of $\gamma$ and cannot explain the non linear behavior of the degree distribution at high values of $k$ (as outlined in \cite{Guimaraes2005}). 
Reproduced in Table \ref{tab:degreedist} from the data in \cite{Albert2002} are some key parameters from a selection of the analyzed real world networks. 
The data is taken from a wide range of sources, which we supplement in Section \ref{sec:analysis}, including the classic movie actor collaboration network from IMDB, a physical communications network, a biological network and a number of collaboration networks.
A striking feature of all of these networks is both a limit to the degree of a node, and also that the value of $\gamma$ is significantly lower than predicted by the scale free model ($\gamma$ is calculated as described in Section \ref{methods}.). Recent work \cite{Berger2014} has highlighted a number of deficiencies in the scale free model, including deviations from the scale free degree distributions and the presence of cut offs in the maximum degree. It must be stated however that the model is strikingly powerful in its ability from a simple set of assumptions to explain many features of complex networks, from their small world property to the absence of a `scale' in the degree distributions. This simplicity is powerful and hints at fundamental processes underlying the dynamics of network evolution.

Failure to capture the detail of the degree distributions of real world networks however, indicates that this simplicity must be supplemented with additional facets to the model of node attachment. In addition the appeal to node degree being the primary determinant of  attachment probability is a modeling assumption and does not explain \emph{why} that is the case. The principal argument is based on the concept of ``the rich get richer'', which is an equivalent statement to equation (\ref{eqn:pref_pi}). In our view this is not a `first principles argument', based upon fundamental physics. Given the success of the model and  widespread acceptance of its validity and application in many fields from genetics to network design, it would be satisfying to link the derivation of equation (\ref{eqn:pref_pi}) to core principles of physics. In this paper we start by exploring a next degree of approximation to the model to identify how environmental influences such as the presence of a top constraint for node degree alter the form of equation (\ref{eqn:pref_pi}). In the model we propose this yields polynomial terms in $k$, which we hypothesize may be part of a series of corrections to the attachment probability. Using arguments based upon applying ensemble statistical mechanics to the entropy of a network vertex, we then propose an entropic model which naturally produces the concept of preferential attachment and constraints, and hints at further structure to the form of attachment probability in equation (\ref{eqn:pref_pi}).

\begin{table}[htbp]
\centering
\caption{Degree Distribution Parameters of some Real Networks \cite{Albert2002}}
\label{tab:degreedist}
\resizebox{\columnwidth}{!}{%
\begin{tabular}{|c|c|c|c|} \hline
\textbf{Source}&\textbf{$\langle k \rangle$}&\textbf{Max Degree}&\textbf{$\gamma$ } \\ \hline
IMDB Movie Actors&28.78/127.33\footnote{The second value of $\langle k \rangle$ is obtained from data in \cite{Herr2007}, and is considered more accurate}&900&2.3 \\ \hline
Internet Router&2.57&30&2.48 \\ \hline
Metabolic, $\emph{E. coli}$&7.4&110&2.2 \\ \hline
Co-authors, SPIRES&173&1100&1.2 \\ \hline
Co-authors, neuro&11.54&400&2.1 \\ \hline
Co-authors, math&3.9&120&2.5 \\ \hline
\end{tabular}
}
\end{table}

\subsection{Extensions to the Scale Free Model}
Before embarking on an investigation of our model, it is important to stress that many proposals to extend preferential attachment have been advanced. These alternative models to preferential attachment rely upon modifications to the probability of attachment beyond simple dependence on the degree of the node. The extensions range from ecologically inspired models such as the competition based approach of D'Souza in \cite{DSouza2007}, to direct alterations of the form of equation (\ref{eqn:pref_pi}) by introducing `super-linear' terms in $k$, that is arbitrary powers of $k$. The model of Krapivsky {\sl et al} \cite{Krapivsky2000}, explicitly explores forms of attachment probability where the term in k is replaced by an exponential form $k^{\alpha}$, where the exponent $\alpha$ can vary in the range $0 < \alpha < \infty$. By varying $\alpha$ it is possible to and produce very different forms of the degree distribution. These range from stretched exponential degree distribution to a super-linear zone for $\alpha > 2$ where one node captures a connection to all other nodes. In other work, notably Dorogovtsev {\sl et al} \cite{Dorogovtsev2000}, the concept of initial attractiveness of a node is introduced, which permits values of the power law exponent to vary and produces values of $\gamma$ that are between $2 < \gamma < 3$. These models depend upon the concept of some nodes starting with a higher initial attractiveness than others in their ability to gain connections to new nodes. In some ways this is the opposite approach to the constrained model we propose in this paper, where nodes become progressively less attractive as they acquire connections and approach their limit.

It is perhaps the ecological, and physically inspired extensions that are most attractive alternatives to preferential attachment. We have already mentioned the competition based model of D'Souza \cite{DSouza2007} that uses an optimization approach in which the minimization of a cost function upon every node addition is used to determine which node the new node attaches to. This model produces an exponentially corrected degree distribution of the form $P(k) \propto k^{-\gamma}e^{-\alpha k} $. This degree distribution is similar to that which we see in the data analyzed in Section \ref{sec:analysis}, and is an encouraging advance on the original preferential attachment model.

Another widely accepted approach, which builds upon the work of Dorogovtsev, was developed by Barab{\'a}si in collaboration with Bianconi, This model parametrizes the attractiveness of the node using a \emph{fitness} measure, $\eta_{i}$, and was introduced in \cite{Bianconi2001}, \cite{Barabasi2000} and further developed in the work of Moriano {\sl et al} \cite{Moriano2013}, and Su {\sl et al} \cite{Su2012}.

The extended model proposes that the probability of attachment is modified to include the fitness parameter in the most general sense, as follows:

\begin{equation}\label{eqn:fitness_attachment}
	\Pi_{i}=\frac{\eta_{i} k_{i}}{ \sum_{j} \eta_{j} k_{j} } \mbox{ .}
\end{equation}

To prevent this model requiring as many independent variables as there are nodes, the attractiveness $\eta$ is fixed, or quenched, at node addition and is randomly assigned from an assumed probability distribution $\rho(\eta)$ for the parameter. The model permits an analogy between the graph and the Bose-Einstein treatment of ideal gases. This analogy relies upon the identification of a node vertex with an energy level of the gas $\epsilon_{i}$, with the degree corresponding to the occupancy number of the energy level. Derivation of graph properties from statistical mechanical arguments is long established, including in the work of Newman and Park on exponential random graphs described in \cite{Park2004}. In the Bianconi-Barab{\'a}si model the fitness parameter is defined as $\epsilon_{i}= - \frac{1}{\beta} \log \eta_{i}$, with $\beta$ being identified as classical inverse thermodynamic temperature. The denominator of equation (\ref{eqn:fitness_attachment}) is then easily identified with the partition function $Z$, familiar from the Bose-Einstein model of statistical mechanics. Using the probability distribution $\rho(\eta)$ of the nodes' fitness parameter as outline in \cite{Bianconi2001}, $P(k)$ can be analytically solved for in the case of the uniform distribution to yield:

\begin{equation}
	P(k) \sim \frac{ k^{-1+C} }{ \log (k) } \mbox{, where $C$ is a constant }
\end{equation}

This model is attractive, and indeed does provide a closer fit to the data, including the presence of a cut-off on the maximum degree of a node.

The models described thus far all share a similar set up to the original preferential attachment mechanism, in that they consider a stepwise addition of a single node which connects to a variable number of pre-existing nodes. In recent work by Bianconi {\sl et al}, this has been generalized to investigate models based upon the addition of simplicial complexes to a network rather than nodes as described in \cite{Bianconi2016,Courtney2017}. These models, referred to as Network Geometry with Flavor (NGF), introduce the concept of a $d$ dimensional simplex, which is a fully connected clique of $d+1$ nodes. When $d=1$ the model reduces down to the Bianconi-Barab{\'a}si model, but higher dimensional simplices are hypothesized to more correctly represent the growth of networks where the unit of addition is a clique, such as a citation network being built from sub networks of frequently collaborating authors. The NGF model proceeds by adding a single node and links, so as to produce a new $d$ dimensional simplex in the graph, by attaching the simplex to a randomly chosen $d-1$ existing face in the graph, governed by a generalized form of equation (\ref{eqn:fitness_attachment}). The attachment probability is further parameterized by a flavor variable $s$ which can take the values of $-1,0,1$ that allows the introduction of a generalized degree which counts the number of $d$ dimensional simplices incident to a node. The range of flavor ensures that the form of attachment probability, which is beyond the scope of this survey to outline, produces a well behaved probability. The survey in \cite{Bianconi2016} has a full and complete overview of the model. The attraction of these models is the generation of a rich set of possible graph geometries, including scale free, Apollonian and a form of graph deeply analogous to the form of graphs proposed in a range of approaches to Quantum Gravity.

Together with the competition model of D'Souza these more physically and ecologically inspired models provide motivation to explore other analogies with such processes to improve upon the standard preferential attachment. It would be a significant insight if we could explain the experimental data based upon solely intrinsic properties of the graph such as node degree and local clustering coefficient of a node, with reference to how these relate to fundamental properties such as entropy and constraints. In the next section we propose  an extension, based upon the concept of constraints to the maximum degree of a node. This constraint is motivated from real world concerns in many networks. For example in communications networks the number of physical connections a node can maintain has a hard limit, and even in social networks building a network of friends is subject to constraints of time and physical space. In Section \ref{sec:dynamics} we show how both constraints and non-linear preferential attachment could arise from a deeper, more fundamental, entropic model.

\section{A Pure Constraint Based Model}
\label{sec:constraints}
A core assumption of the scale free model is that new nodes attach to other nodes with a probability that is determined only by the degree of the target node; no other factors affect $\Pi_{i}$ and attachment is unconditional. In most networks though this is not a fully accurate assumption, as most nodes will have some inherent upper limit on their capability to establish connections. We can imagine a network comprised of nodes capable of maintaining a maximum of $c$ connections, with $c_{i}(t)$ being the point in time capacity of node $i$ at time $t$. To simplify the treatment we assume the capacity of all nodes is equal across the network. In this case we could imagine modifying the probability of attachment to account for the nodes capacity as they accumulate connections, with a multiplicative factor to the preferential attachment probability $\Pi_{i}$. This assumption of uniform maximum capacity is an approximation that we justify by the simplicity of the theoretical analysis it permits. We seek to avoid introducing a family of free parameters, which would equate to a family of constraints, to preserve the theoretical elegance of the treatment. When we come to compare our \emph{constrained} model to real world data it does require us to make reasonable estimates for the effective average constraint.
We assume that this acts as a scaling factor for the attachment probability, similarly to the fitness factor introduced in the Barab{\'a}si-Bianconi model \cite{Bianconi2001},\cite{Barabasi2000}, in essence acting like a conditioning of the probability of attachment with the probability the node can accept the connection. 
In the most general sense, we can write this as the ratio of the nodes capacity relative to the time varying, average capacity of an arbitrary node,  $\langle c(t) \rangle$  as:

\begin{equation}\label{eqn:bare_prob}
\begin{split}
	\Pi_{i}^{c}=\zeta_{i}  \times \Pi_{i} \mbox{, where } \zeta_{i}=\frac{(c-k_{i}(t))}{\langle c(t) \rangle}\\
	 \mbox{ and } \Pi_{i} =\frac{k_{i}(t)}{2mt}
\end{split}
\end{equation}

To calculate $\langle c(t) \rangle$, we  observe that at any time $t$ a given node $i$ will have an expected value of capacity $\langle c_{i}(t) \rangle = \langle c-k_{i}(t) \rangle$. As we assume that $c$ is a shared maximum capacity across all nodes this reduces to $\langle c_{i}(t) \rangle = c-\langle k_{i}(t) \rangle$, and we note that $\langle k_{i}(t) \rangle$ is the expected value of a node's degree $k_{i}=\langle k_{i} \rangle$, which will be useful in section \ref{sec:analysis} when we will compare our constrained model against real networks. We can also estimate the expected value of the capacity of a node, by assuming a base uniform distribution of attachments in the absence of preference. After $n$ nodes have been added, we will have added $nc$ capacity to the graph, and consumed $2nm$ connections. In the simplest case for the average capacity of a node, after adding a large number of nodes $n$, we note that the average capacity must evolve to a constant as following:

\begin{equation}\label{eqn:avgcap}
	\langle c(t) \rangle = \frac{nc-n2m}{n}=c-2m \mbox{ .}
\end{equation}

Unfortunately as written this attachment probability is not sufficient as $\sum\limits_{i} \Pi_{i}^{c} \neq 1$. This can be demonstrated by expanding Equation (\ref{eqn:bare_prob}) as follows:

\begin{equation*}
\begin{split}
	\sum\limits_{i} \Pi^{c}_{i} = \frac{ 1 } {(c-2m)2mt} \sum\limits_{i} (c-k_{i}(t))k_{i}(t) \mbox{, }\\
		= \frac{ 1 } {(c-2m)2mt} \Bigg\{ c2mt - \sum\limits_{i} k_{i}(t)^{2} \Bigg \} \mbox{ .}
\end{split}
\end{equation*}

If we define $\delta$ as 

\begin{equation}\label{eqn:delta}
	\delta=\frac{ \sigma } { \sum\limits_{i} k_{i}(t) } = \frac{\sigma}{ 2mt } \mbox{ where, }
\end{equation}
\begin{equation}\label{eqn:sigma_num}
	\sigma=\sum\limits_{i} k_{i}(t)^{2} - \sum\limits_{i} \langle k_{i}(t) \rangle^{2}
\end{equation}

the normalization sum becomes,

\begin{equation*}
	\sum\limits_{i} \Pi^{c}_{i} = 1 - \frac{ \delta }{ c-2m }\mbox{ . }
\end{equation*}

In general $\delta$ could be a function of time and degree, but as an approximation in our model we treat it as a constant of the system. We test that assumption in the simulations presented later in this section, which indicate that it is valid to assume that $\delta$ eventually stabilizes to a constant as the network evolves. We run these simulations of network growth to mimic the parameters for a selection of the real network data we analyze. Investigation of models where $\delta$ is a function of time (and potentially $k_{i}$) is an current avenue of research, and the subject of future work.
For our attachment probability to be a valid probability measure we need to establish that $\frac{\delta}{(c-2m)} \geq 0$ and that $\frac{\delta}{(c-2m)} \leq 1$. In the first instance the numerator of Equation (\ref{eqn:delta}), as defined in Equation (\ref{eqn:sigma_num}),  is the variance of $k_{i}$ across the graph, and so is strictly positive. Providing that $c > 2m$, we can safely assume $\delta \geq 0$.

Regarding the upper limit of $\delta$, we can appeal to Popviciu's inequality (see \cite{Sharma2014}) for a bounded distribution, with $k_{max} = c$ and $k_{min}=m$. This states:

\begin{equation*}
\begin{split}
	\sigma \leq \frac{1}{4} (k_{max} - k_{min} )^2 \leq \frac {1}{4} (c-m)^2 \mbox{ ,}\\
		\Rightarrow \frac{\delta}{(c-2m)} \leq \frac{ (c-m)^{2} }{ 8(c-2m)mt }	\mbox{ .} 
\end{split}
\end{equation*}

For times $t > \frac{ (c-m)^2 }{8m(c-2m) }$, we then conclude that as required $\frac{\delta}{(c-2m)} \leq 1$. With these limits established, we can modify the attachment probability by adding in $\delta$ to produce a  form for the attachment probability, which sums to unity at each time step across all nodes, below:

\begin{equation}\label{eqn:norm_prob}
\begin{split}
	\Pi_{i}^{c}=\zeta_{i}  \times \Pi_{i} \mbox{, where } \zeta_{i}=\frac{(c+\delta-k_{i}(t))}{c-2m}\\
	 \mbox{ and } \Pi_{i} =\frac{k_{i}(t)}{2mt}
\end{split}
\end{equation}

For convenience, we can further simplify the  expression for $\zeta_{i}$, as follows:

\begin{equation}
\begin{split}
	\zeta_{i}=\alpha \Bigg(1-\frac{k_{i}(t)}{(c+\delta)} \Bigg) \mbox{,}\\
	 \mbox{ where } \alpha=\frac{c+\delta}{c-2m} \mbox{, or equivalently } \alpha=\frac{c+\delta}{c-2 \langle k_{i} \rangle} \mbox{ .}
\end{split}
\end{equation}

We can now write the complete probability of attachment as:

\begin{equation}\label{eqn:constr_prob}
	\Pi_{i}^{c}=\frac{ \alpha k_{i}(t)(c+\delta-k_{i}(t)) } { 2m(c+\delta)t } \mbox{ .}
\end{equation}
\\
For comparison with the  Barab{\'a}si-Albert model, using $\alpha=\frac{c+\delta}{(c-2m)}$ from equation (\ref{eqn:avgcap}) we can rewrite $\Pi_{i}^{c}$ as follows:

\begin{equation*}
	\Pi_{i}^{c}=k_{i}(t) \frac{ (\frac{c+\delta-k_{i}(t)}{c-2m}) }{2mt} \approx k_{i}(t) \frac{1}{2mt} \mbox{, for large $c$.}
\end{equation*}
\\
This recovers the standard Barab{\'a}si-Albert model in the case that the constraint $c$ is infinite and therefore does not interfere with the dynamics of the network's evolution.

Following the continuum approach, and dropping the explicit time dependency of $k_{i}$ for clarity, we can substitute this into equation (\ref{eqn:pref_attachment}), to obtain

\begin{equation}
	\frac{\partial{k_{i}}}{\partial{t}}=m\Pi_{i}^{c}=\frac{ \alpha k_{i}(c+\delta-k_{i}) } { 2(c+\delta)t } = \frac{\alpha k_{i}}{2t} - \frac{\alpha k_{i}^{2}}{2(c+\delta)t} \mbox{ ,}
\end{equation}

with the fraction multiplied out for convenience later. This is directly solvable by separating as follows:

\begin{equation*}
\begin{split}
	\frac{1}{\alpha} \int \frac{ dk_{i} }{k_{i}(c+\delta-k_{i})} = \frac{1}{\alpha (c+\delta)} \int \Big\{ \frac{1}{k_{i}} + \frac{1}{c-k_{i}}  \Big\} dk_{i}\\
	 = \frac{1}{2(c+\delta)} \int \frac {dt}{t} \mbox{ ,}
\end{split}
\end{equation*}

whose solution is:

\begin{equation*}
\begin{split}
	\log \Bigg( \frac{k_{i}}{c+\delta-k_{i}} \Bigg) = \frac{\alpha}{2} \log (t) + \theta \mbox{ ,}
\end{split}
\end{equation*}

or in simplified form

\begin{equation*}
	k_{i}=(c+\delta) e^{\theta} \Bigg( \frac{ t^{\alpha/2} } { e^{\theta}  t^{\alpha/2} +1 } \Bigg) \mbox{ .}
\end{equation*}

Following the continuum method in \cite{Albert2002} we apply the initial condition that $k_{i}(t)=m$ at time $t=t_{i}$, to obtain:

\begin{equation}
\begin{split}
	k_{i}(t)=\Bigg(  \frac{ \rho (c+\delta) \big( \frac{t}{t_{i}} \big)^{\alpha/2}) }{ 1+ \rho \big( \frac{t}{t_{i}} \big)^{\alpha/2} } \Bigg), \\
	\mbox{ with $\rho$ defined as, } \rho=\frac{m}{c+\delta-m} \mbox{ .}
\end{split}
\end{equation}

Again, we note that as $c \rightarrow \infty$, $\rho (c+\delta) \rightarrow m$, $\alpha \rightarrow 1$, and so equation (6) reduces to

\begin{equation*}
	k_{i}(t)=m \Bigg(  \frac{t}{t_{i}}\Bigg)^{1/2} \mbox{ ,}
\end{equation*}

the standard result from the continuum analysis of Barab{\'a}si and Albert \cite{Albert2002},\cite{Albert1999}. We then note that the probability that a node has degree $k_{i}(t) < k$ is:

\begin{equation*}
\begin{split}
	P(k_{i}(t) < k)=P\Big( t_{i} > \frac{ \rho^{2/\alpha}(c+\delta-k)^{2/\alpha}t }{k^{2/\alpha}} \Big) \\
	= 1 - P\Big( t_{i} \leq \frac{ \rho^{2/\alpha}(c+\delta-k)^{2/\alpha}t }{k^{2/\alpha}} \Big) \mbox{ .}
\end{split}
\end{equation*}

Assuming uniform probability for the choice of node introduction time $t_{i}$ of $ \frac{1}{(m_{0} + t)}$ we arrive at the expression:

\begin{equation*}
	P(k_{i}(t) < t) = 1 - \frac{ \rho^{2/\alpha}(c+\delta-k)^{2/\alpha}t }{k^{2/\alpha}(m_{0} + t)} \mbox{ .}
\end{equation*}

Although somewhat more complex than the expression in \cite{Albert2002} it is nevertheless simple to compute the distribution equation $P(k)=\frac {\partial{ (k_{i}(t) < k) }} {\partial{k}}$ to obtain the main result of our constrained model:

\begin{equation}\label{eqn:main_result}
	P(k) = \frac{2 (c+\delta)\rho^{2/\alpha}t}{\alpha(t+m_{0})} \Bigg(  \frac{ (c+\delta-k)^{\frac{2}{\alpha}-1}} {k^{ \frac{2}{\alpha}+1}} \Bigg)  \mbox{ .}
\end{equation}

In appendix \ref{app:gamma_deriv} we examine the asymptotic behavior of Equation (\ref{eqn:main_result}), which verifies that by careful manipulation  the standard result of the scale free model $\gamma=3$, is recovered in the limit $c \rightarrow \infty$. Further, this analysis also indicates that the dominant contribution to degree  distribution for $k \ll (c+\delta)$, produces a scale free log linearity with power law exponent $\gamma=\frac{2}{\alpha} + 1$. This equivalence to a more straight forward power law, but with an exponent $\gamma < 3$ for values of $k \ll (c+\delta)$ indicates that the presence of a constraint influences the behavior of our model even for nodes early in their evolution. This is a significant result and we make use of it to compare the predictions of our theory against real network data and simulations in section \ref{sec:analysis}.

\begin{figure*}[t]
	\centering
	\begin{subfigure}[t]{0.45\textwidth}
		\centering
		\includegraphics[scale=0.4]{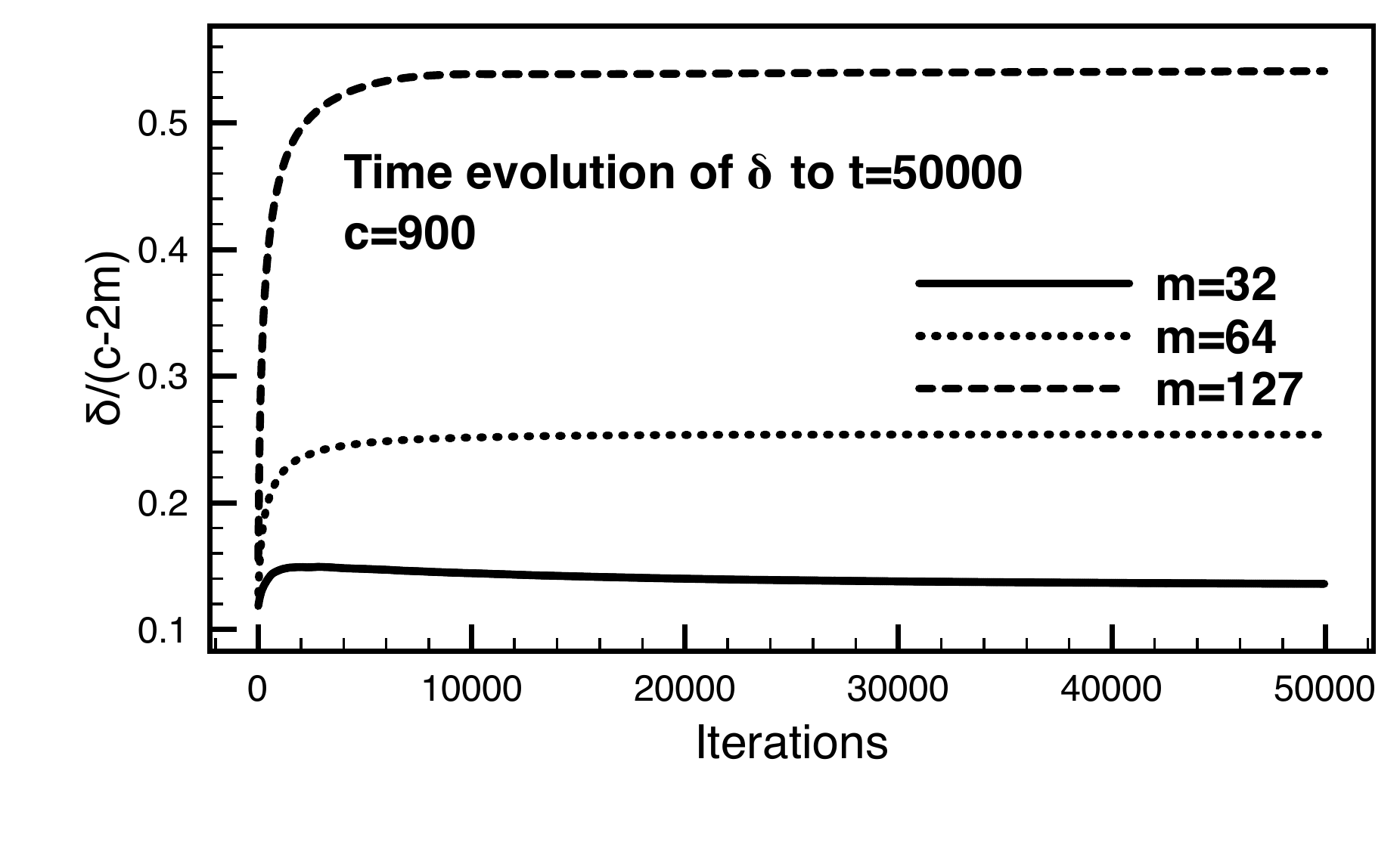}
		\caption{Convergence of $\frac{\delta}{(c-2m)}$ over $50,000$ Iterations in a Simulation of Constrained Attachment}
		\label{fig:convergence}
	\end{subfigure}%
	~ 
	\begin{subfigure}[t]{0.45\textwidth}
		\centering
		\includegraphics[scale=0.4]{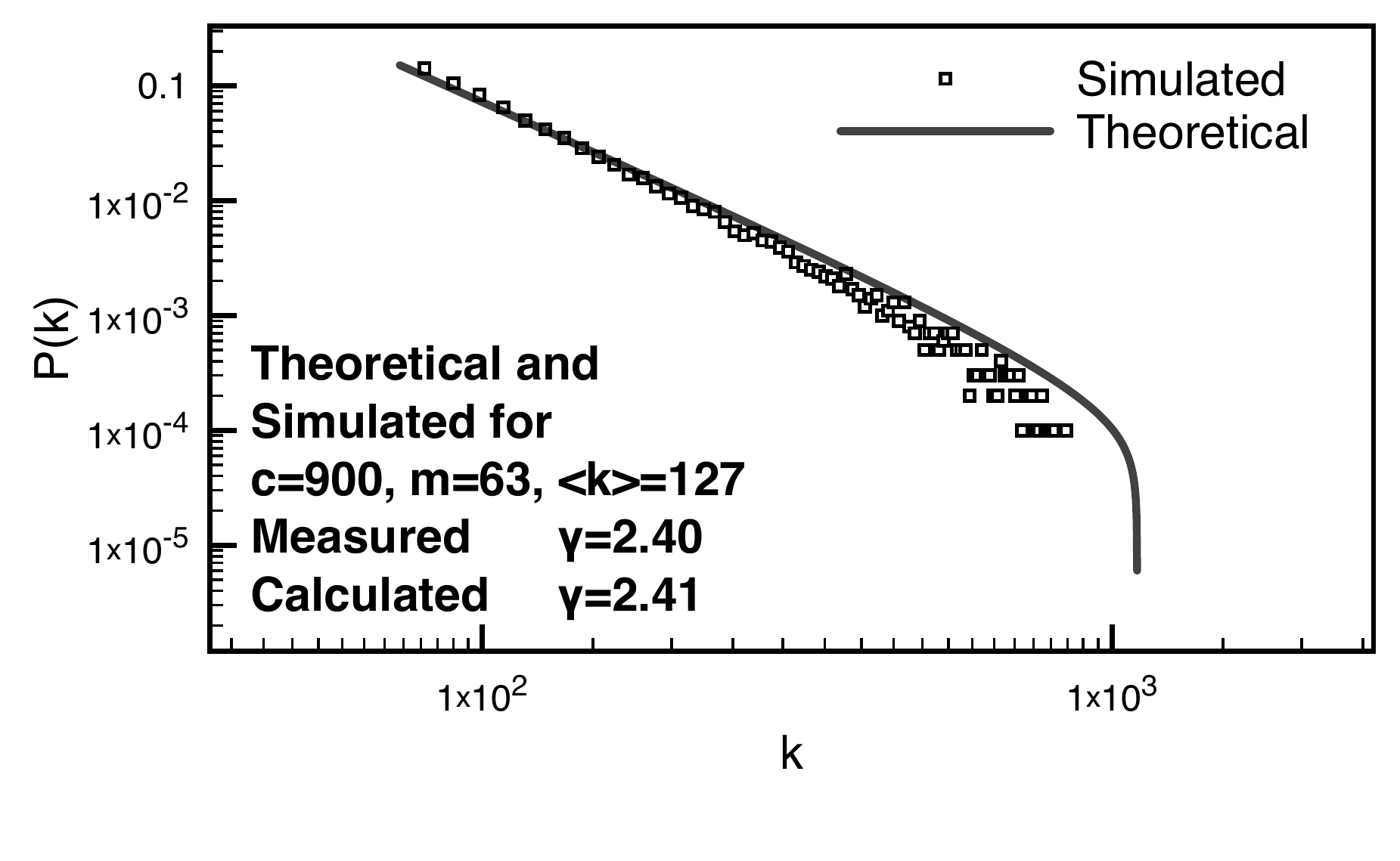}
		\caption{Simulation and Theoretical  Degree Distribution using Equation(\ref{eqn:constr_prob}) and IMDB Parameters at $t=50,000$}
		\label{fig:imdb_simul}
	\end{subfigure}
	~
	\begin{subfigure}[t]{0.45\textwidth}
		\centering
		\includegraphics[scale=0.4]{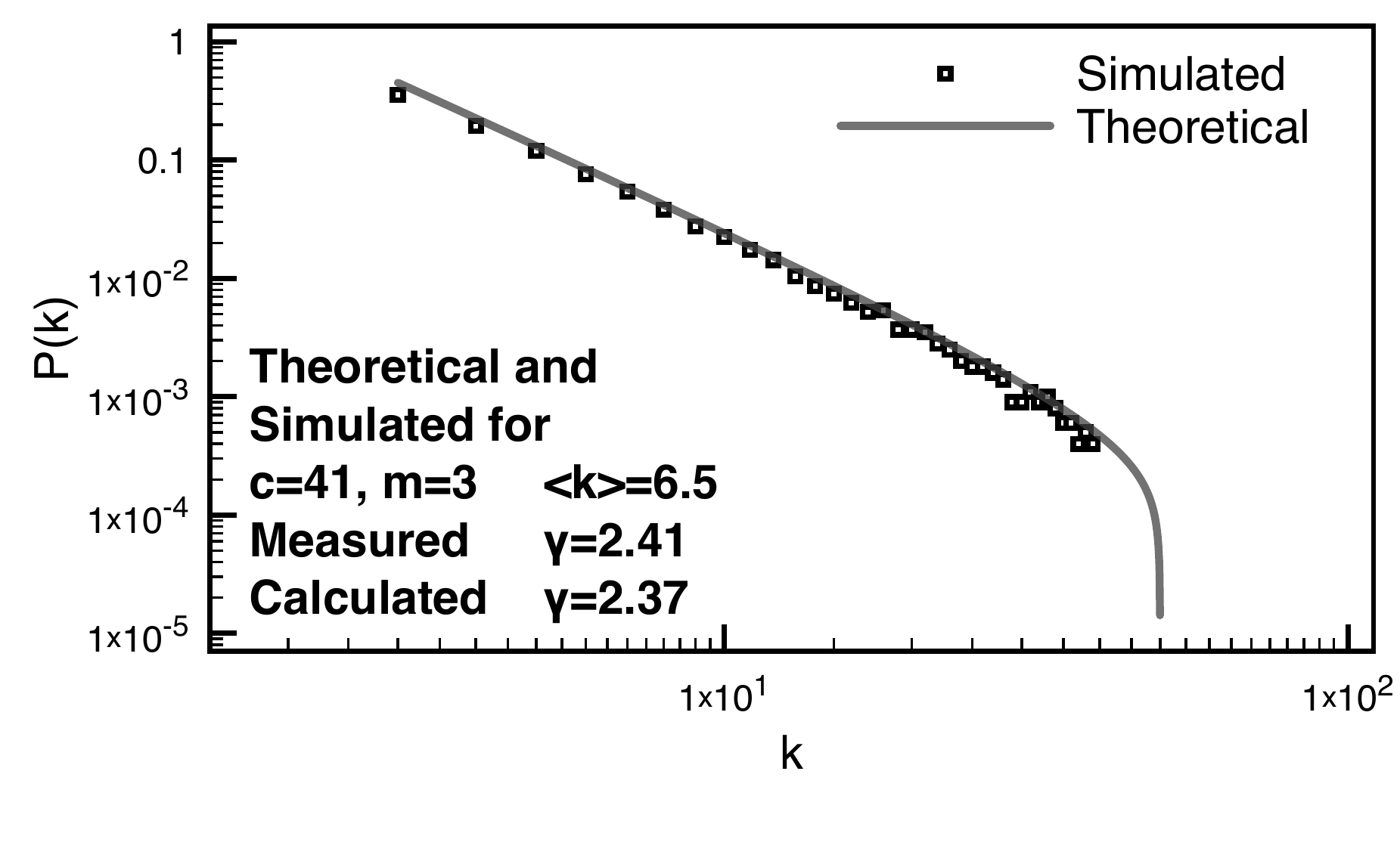}
		\caption{Simulation and Theoretical  Degree Distribution using Equation(\ref{eqn:constr_prob}) and Patents Parameters at $t=50,000$}
		\label{fig:patent_simul}
	\end{subfigure}%
	~ 
	\begin{subfigure}[t]{0.45\textwidth}
		\centering
		\includegraphics[scale=0.4]{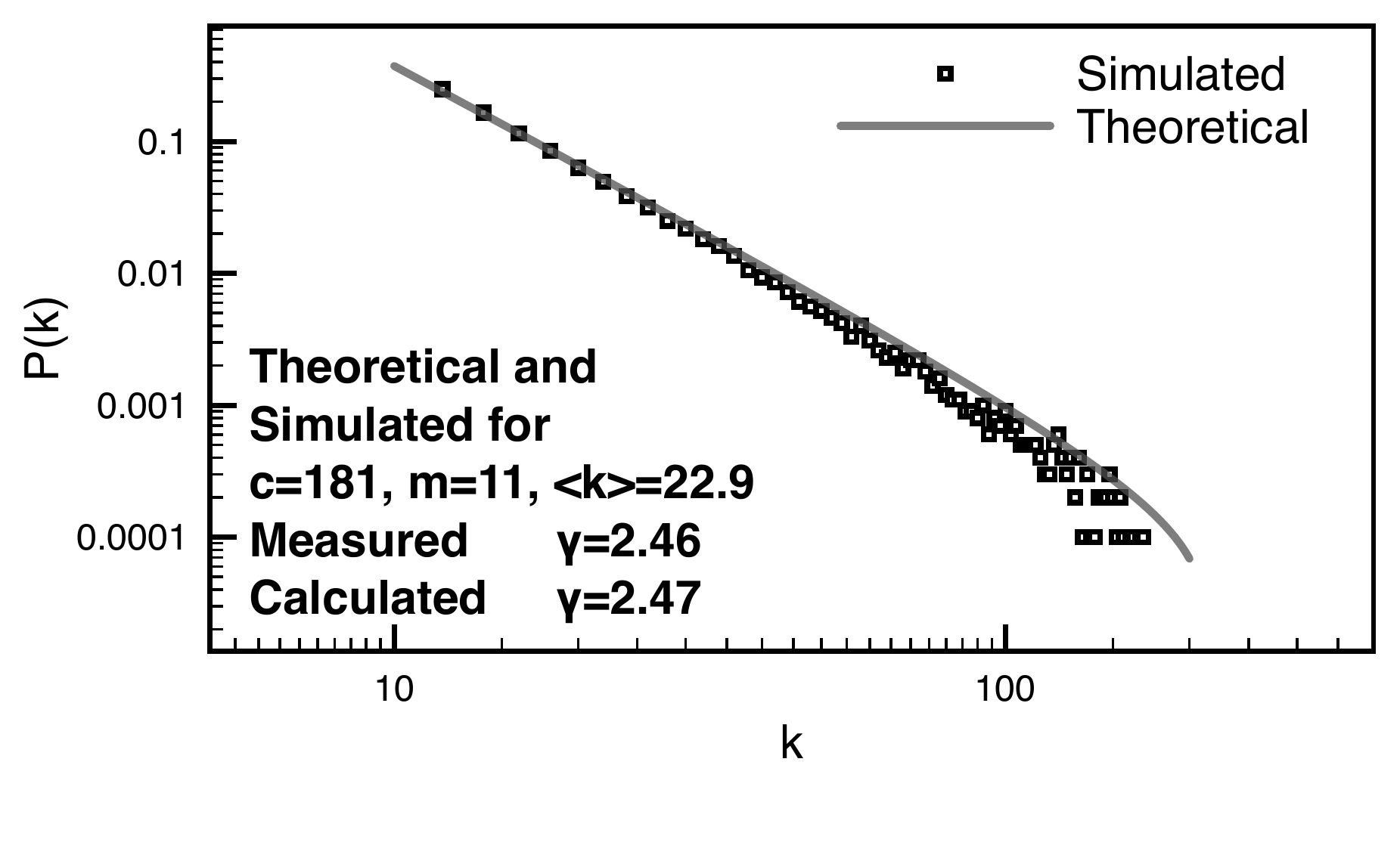}
		\caption{Simulation and Theoretical  Degree Distribution using Equation(\ref{eqn:constr_prob}) and Web Provider Parameters at $t=50,000$}
		\label{fig:yahoo_simul}
	\end{subfigure}
	\caption{Simulation Results for Constrained Attachment}
	\vspace{-1.5em}
\end{figure*}

The result in equation (\ref{eqn:main_result}) has some interesting implications, as the presence of a finite capacity $c$ alters the scale factor for the distribution of the nodes, \emph{whilst} preserving the essential aspects of scale free behavior. 
By way of example, the data for the IMDB movie actor database, as presented in Table \ref{tab:degreedist}, is plotted in Figure \ref{fig:imdb_simul}, along with results from a simulation of our model. The movie actor database naturally produces a graph by assigning a vertex for each actor and connecting two vertices when the actors have acted in the same film. Figure \ref{fig:imdb_simul} contains a theoretical plot of the distribution taken directly from equation (\ref{eqn:main_result}), using $\langle k \rangle=127$, $c=900$ and with initial conditions of $m_{0}=100$, which we take from Table \ref{tab:degreedist} . For this plot we set $\delta=205$, which we take directly from the simulation, which we discuss in the next paragraph.
The unmodified scale free model would give a value of $\gamma$ of exactly 3, but our modification has an initial value of $\gamma=\frac{2}{\alpha}+1$, which increases as $k \rightarrow c$ and reaches a limit when $k=c$. 
To calculate $\gamma$ we can take $c=900$ from the dataset in Table \ref{tab:degreedist} and $k=127.33$, with the estimated value of $\delta=243$ (we average the ratio of $\delta$ to $c$), to yield $\gamma=2.35$, versus the measured value of 2.3 in \cite{Albert2002} and 2.43 from our simulation. By comparison, to the scale free model, our approach predicts the value of $\gamma$ to 2.29\%, compared to  30.4\% for scale free, a significant improvement. In addition, there is no explanation in the scale free model for the degree of a node in the graph having a maximum value.

To further verify our model, and in particular the assumption that $\delta$ can be effectively treated as a constant, simulations were run using the form of preferential attachment probability in equation~(\ref{eqn:constr_prob}), for a network sharing the same parameters of maximum degree and average degree as the IMDB network. We present those results in Figure \ref{fig:convergence}. The simulation was run for a selection of initial parameters to asses the evolution of $\delta$, and in each case the value quickly converges to a constant. Turning to the simulation of degree distribution, in Figure \ref{fig:imdb_simul} the essential scale free nature of the network obtained is visible on the log scale graph, as is the goodness of fit and agreement between the simulation with a theoretical plot of $P(k)$ using the same simulation parameters. Using the techniques described in \cite{Clauset2009}, we can measure $\gamma$, and obtain a value of $2.40$ versus a calculated value from equation~(\ref{eqn:main_result}) of $2.41$, which is in close agreement.

We also ran simulations for the Patents Citation graph (Figure \ref{fig:patent_simul}) and the Web Provider network (Figure \ref{fig:yahoo_simul}), which both produce similarly good results to the IMDB network in terms of the closeness of fit between the simulated and theoretically obtained $P(k)$. We can conclude that the constrained model is a good representation of networks with a simple maximum degree constraint.

Motivated by this example and simulation, in the following section we extend our analysis to a range of more recent, publicly available, network data to investigate further the accuracy of our constrained model.

\section{Analysis and Comparison of Constrained versus Preferential Attachment}
\label{sec:analysis}

\subsection{Data and Methods}
\label{methods}

In this section we present the analysis of an extensive collection of network datasets comprising virtual, transport, and communications networks. The bulk of this data is publicly available through the Stanford Large Datasets Collection \cite{snapnets} which comprises an excellent repository of large graphs.  The Twitter follower data is provided by \cite{Cha2010}, and the rest of the datasets are reproduced from publications such as \cite{Albert2002}, the Internet Topology Zoo \cite{Knight2011a}. We have one proprietary graph built from the topology taken from a large commercial deployment of network infrastructure used to deliver a top 10 Internet portal service (see \cite{Tee2016b}).

The produced graphs fall into the following categories:

\begin{itemize}
    \item \emph{Social Networks}. These include Twitter, Facebook, Pokec graphs of the relationships between users. Typically each user is a node and nodes have links if the users have some form of relationship with each other. For example in the case of Twitter this relationship derives from one user `following' another.
    \item \emph{Collaboration and Citation Networks}. These cover a wide range of publicly available data, including the Arxiv citation, Patent Citation and co-authorship graphs as examples. Graphs are constructed by creating a vertex for each unique user or paper and then connecting the vertices if they share authorship with another vertex or directly cite it.
    \item \emph{Communications Networks}. These networks, such as the Internet Router, IT Zoo, Web Provider and Berkeley Stanford Web Graph are constructed by representing physical or virtual nodes by a vertex in the graph and communications links as edges connecting the vertices.
    \item \emph{Biological Networks}. These networks use a graph to represent a biological process, for example the metabolism of the \emph{E. coli} organism. Nodes in the  graph represent a molecule or intermediate state in the process used by \emph{E. coli} to release energy from its food sources, with edges connecting nodes where a reaction or transition occurs. Similar networks exist for other biological processes (e.g. for the genetic cause and effect in cancers and disease epidemic spreads).
\end{itemize}

\begin{figure}[t]
\centering
	\includegraphics[scale=0.4]{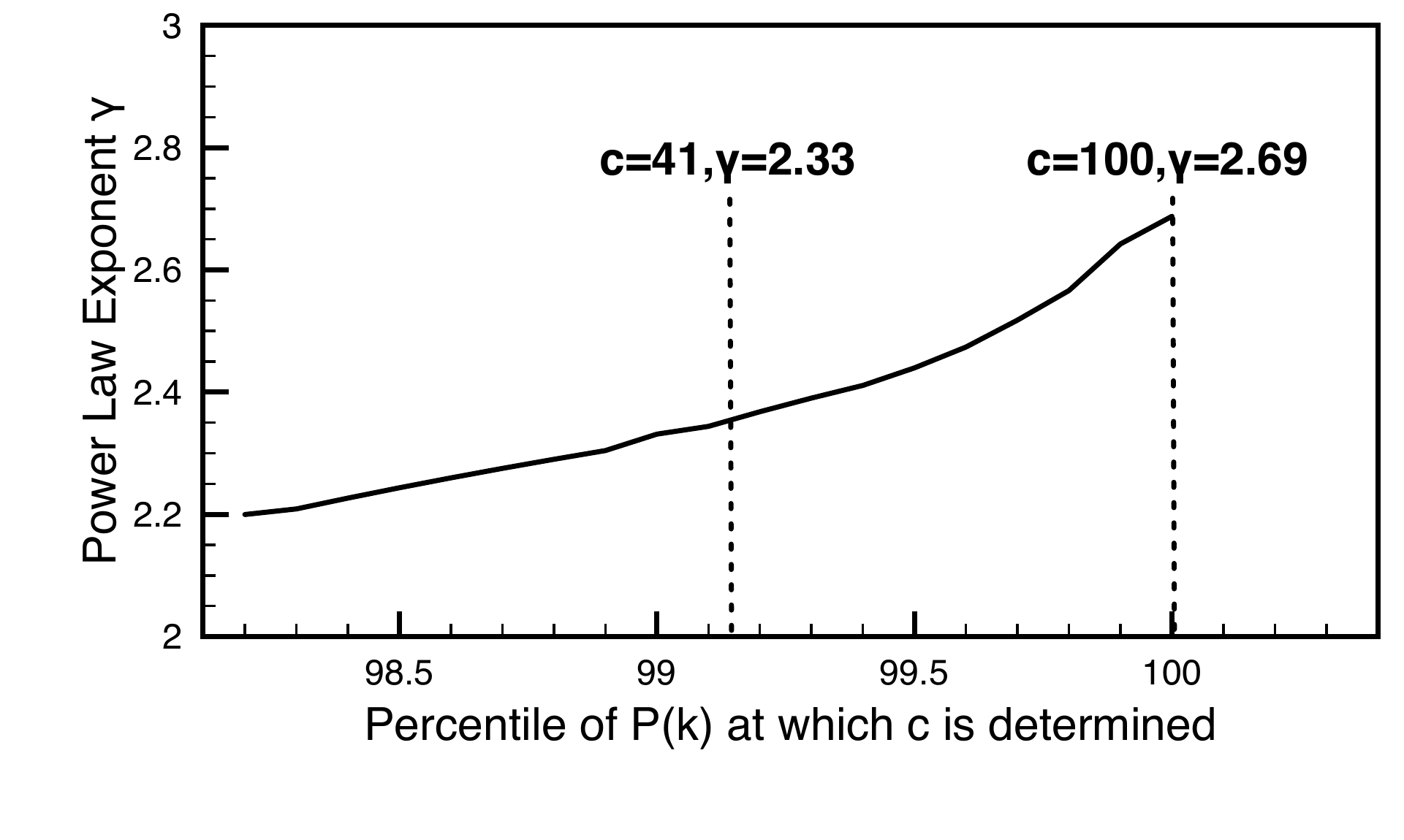}
	\caption{Variation of Calculated Values of $\gamma$ with Choice of Percentile for $c$ for the Patents Graph}
	\label{fig:gamma_var}
\end{figure}

Analysis of the data was undertaken using a program and graph datastore which is available from the authors on request. The source data was often very large (the Twitter data contains for example over 10 million edges), and extracting values for the max degree and $\langle k \rangle$ is not necessarily evident. Some of the data had some extreme outliers in terms of node degree, and to avoid skewing the results, we estimated the constraint at the $99^{th}$ percentile of $k$ rather than the maximum value in the data. This is consistent with the methodology taken in the theoretical analysis, where we made an assumption of the node degree constraint being constant for all nodes. This is a simplification, but one with great benefit in the analytical treatment of the model. The elimination of outliers at first sight may seem inconsistent with the assumption of a single constraint in the capacity of a node, but it is expected that the real world data will contain perhaps many different constraints, and that the average behavior of the graph will be most influenced by the effective maximum established at the $99^{th}$ percentile.
Further, the data above the $99^{th}$ percentile in $k$ is typically very sparse and may contain spurious data points, which this cut off eliminates. In Figure \ref{fig:gamma_var} we present the variation of the calculated value of $\gamma$ with the choice of percentile at which to choose $c$. The range of calculated values as we move from the $98.2^{th}$ to the $100^{th}$ percentile is $2.20$ to $2.69$, a range of $\pm 9\%$ either side of the chosen value of $c=41$. We believe this further strengthens our choice of the $99^{th}$ percentile as the appropriate cut off for measuring $c$. 

For $\langle k \rangle$ we require the expected value of the degree. This was calculated by computing the weighted mean, a discrete approximation of $\langle k \rangle$, which is truly only valid if $k$ is a continuous variable. This is consistent with the approximation of continuity inherent in the  continuum analysis approach.

To compare against the actual value of $\gamma$, power law exponent, we followed the techniques outlined in \cite{Clauset2009} to both asses the presence of a scale free distribution and obtain the value of $\gamma$. For the datasets we analyzed, which can be seen visually in Figures \ref{fig:social-degree}, \ref{fig:citation-degree} and \ref{fig:infra-degree}, there is a considerable portion of the distribution which has a well defined straight line on the log/log plots, illustrating the intrinsic power law distribution of node degree. We capture the measured values of these power law exponents in Table \ref{tab:gammatab}.

\begin{table*}[htbp]
\caption{Comparison of $\gamma$ Predictions Between Preferential Attachment and Constraints Model}
\label{tab:gammatab}
\resizebox{\textwidth}{!}{%
\begin{tabular}{|l|c|c|c|c|c|c|} \hline
	\textbf{Source}&\textbf{$<k>$}&\textbf{$c$}&$\gamma$ Calculated&$\gamma$ Measured&\textbf{$\Delta$ Constraints}&$\Delta$ Scale Free  \\ \hline
	\textbf{Patent Citation}\footnotemark[2]			& 6.57    & 41   &2.33  & 2.31 & \textbf{0.75\%}  & 29.66\% \\ \hline
	\textbf{IT Zoo}\footnotemark[1]					& 2.26    & 10   & 2.32 & 2.36 & \textbf{1.63\%}  & 27.19\% \\ \hline
	\textbf{Internet Router}\footnotemark[7]        		& 2.57    & 30   &2.44 & 2.48 &	\textbf{1.64\%}  & 20.97\% \\ \hline
	\textbf{Arxiv - Condensed Matter}\footnotemark[2]	& 9.13    & 51   &2.32 & 2.37 & \textbf{2.18\%}  & 26.39\% \\ \hline
	\textbf{IMDB Movie Actors	}\footnotemark[7]         	& 127.33& 900 &2.35 & 2.30 &	\textbf{2.29\%}  & 30.43\% \\ \hline
	\textbf{Pokec}\footnotemark[3]					& 39.27  & 180 &2.30 & 2.25 & \textbf{2.29\%}  & 33.34\% \\ \hline
	\textbf{Airport Connections}\footnotemark[9]		& 11.18  & 126 &2.35 & 2.29 &  \textbf{2.82\%}	& 31.19\%\\ \hline
	\textbf{Arxiv - HepTh (Cit)}	\footnotemark[2]		& 26.75  & 165 &2.34 & 2.44 & \textbf{3.82\%}  & 23.10\% \\ \hline
	\textbf{Twitter (Circles)}\footnotemark[4]			& 33.94  & 264 &2.37 & 2.47 & \textbf{3.90\%}  & 21.43\% \\ \hline
	\textbf{Arxiv - HepTh (Collab)}\footnotemark[2]	    	& 22.05  & 285 &2.37 & 2.51 & \textbf{5.37\%}  & 19.67\% \\ \hline
	\textbf{Web Provider}\footnotemark[5]			& 4.18    & 36   &2.09 & 2.23 & \textbf{6.33\%}  & 34.48\% \\ \hline
	\textbf{Co-authors, math}\footnotemark[7] 	                	& 3.90    & 400 &2.69 & 2.5   & \textbf{7.60\%}    & 20.00\% \\ \hline
	Berkeley Stanford Web\footnotemark[6]	    		& 24.59  & 173 &2.31 & 2.35 & 10.45\%  		& 27.74\% \\ \hline
	Metabolic, \emph{E. coli}\footnotemark[7]	        		& 53.51  & 137 &2.47 & 2.20 & 12.20\%		& 36.36\% \\ \hline
	AS Skitter\footnotemark[2]						& 54.13  & 150 &1.94 & 2.34 & 17.27\% 		& 28.14\% \\ \hline
	Facebook\footnotemark[4]				            	& 42.99  & 198 &2.25 & 2.75 & 18.08\% 		& 9.23\% \\ \hline
	Arxiv - Astro Phys\footnotemark[2]		    		& 23.81  & 144 &2.33 & 2.87 & 18.70\%  		& 4.61\% \\ \hline
	Co-authors, neuro\footnotemark[7] 	                		& 11.54  & 400 &2.53 & 2.1  &	20.42\% 		& 42.86\%\\ \hline			
	Enron Email\footnotemark[6] 			           	& 40.25  & 280 &1.84 & 2.42 & 23.87\% 		& 24.02\% \\ \hline
	Twitter (Follower)\footnotemark[8]		            	& 8.63    & 90   &1.56 & 2.39 & 34.77\% 		& 25.46\% \\ \hline
	PA Road Network\footnotemark[6]		    		& 5.41    & 9 &1.71 & 2.69     & 36.27\%  		& 11.71\% \\ \hline
	Co-authors, SPIRES	\footnotemark[7]                 	& 173.00 & 1100 &2.69 & 1.2  &	124.17\%		& 150.00\% \\ \hline
	\end{tabular}
}
\footnotetext[1]{Data published in \cite{Knight2011a}}
\footnotetext[2]{Data available at \cite{Leskovec2007}}
\footnotetext[3]{Data published in \cite{Takac2012}}
\footnotetext[4]{Data available at \cite{Leskovec2012}}
\footnotetext[5]{Data analyzed in \cite{Tee2016b}}
\footnotetext[6]{Data available at \cite{Leskovec2011}}
\footnotetext[7]{Data reproduced from \cite{Albert2002}}
\footnotetext[8]{Data supplied by, and, described in \cite{Cha2010}}
\footnotetext[9]{Data available at \cite{Patokallio2016}}
\end{table*}

\subsection{Analysis}
In the summary Table \ref{tab:gammatab} it is compelling to note that in all but a few cases the constrained model is  more accurate in its predictions of $\gamma$ than the standard scale free model. Indeed in the case of the Patent Citation, Internet Topology Zoo, Pokec, the real world network from a Web Provider, and a number of the citation networks and social networks, it comes very close to an exact prediction. Given that the motivation to investigate the constrained model originated from considerations of network design in communications networks, it is interesting to see that this has some strong applicability to non-physical networks.

\begin{figure*}[t]
	\centering
	\begin{subfigure}[t]{0.45\textwidth}
		\centering
		\includegraphics[scale=0.4]{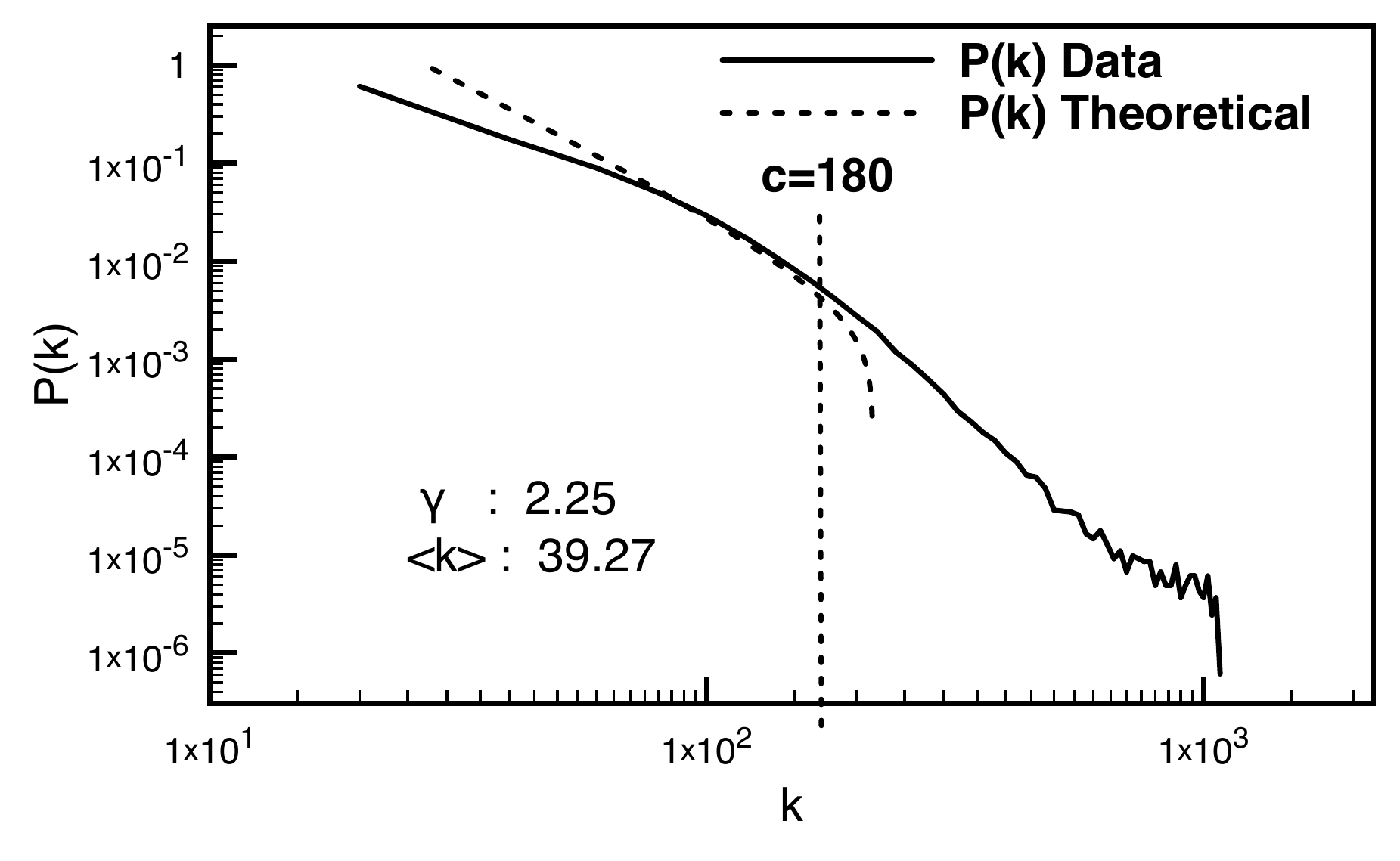}
		\caption{Pokec - Slovakian Social network Friendship Graph, Theoretical and Experimental \cite{Takac2012}}
		\label{fig:pokec}
	\end{subfigure}%
	~ 
	\begin{subfigure}[t]{0.45\textwidth}
		\centering
		\includegraphics[scale=0.4]{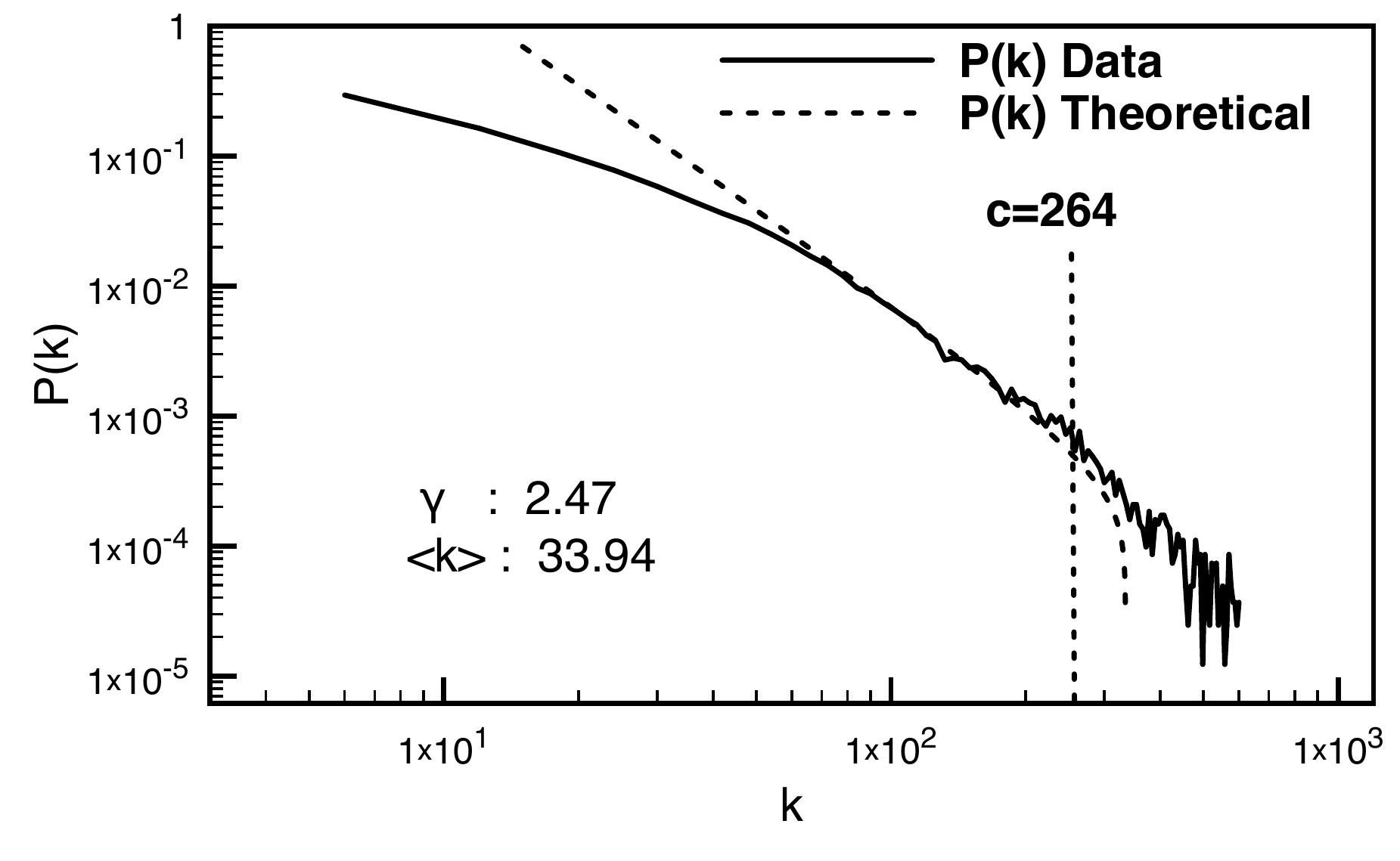}
		\caption{Twitter Friendship Circles, Theoretical and Experimental \cite{Leskovec2012}}
		\label{fig:twittercircles}
	\end{subfigure}
	~ 
	\begin{subfigure}[t]{0.45\textwidth}
		\centering
		\includegraphics[scale=0.4]{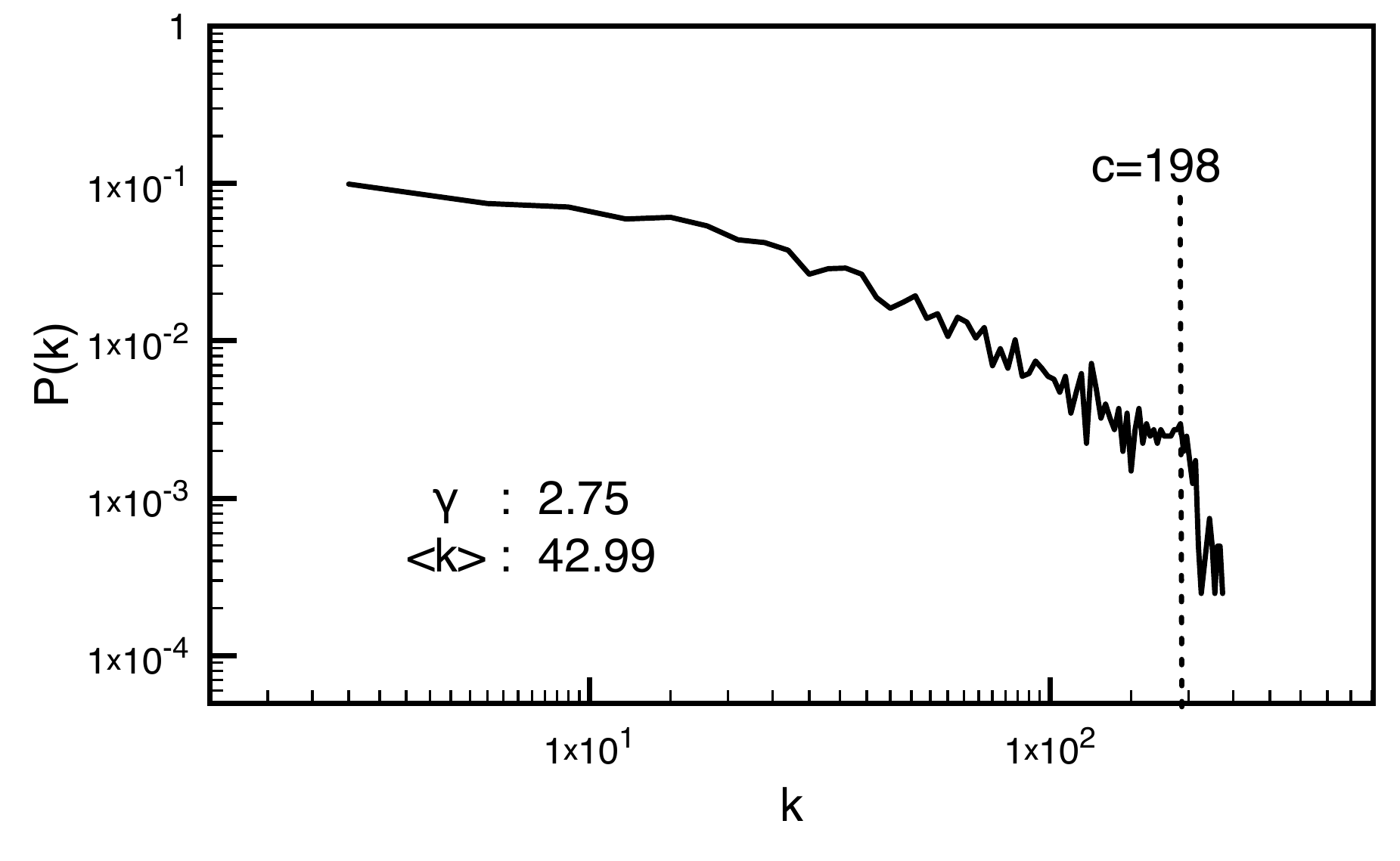}
		\caption{Facebook Friendship Network \cite{Leskovec2012}}
		\label{fig:facebook}
	\end{subfigure}
	~ 
	\begin{subfigure}[t]{0.45\textwidth}
		\centering
		\includegraphics[scale=0.4]{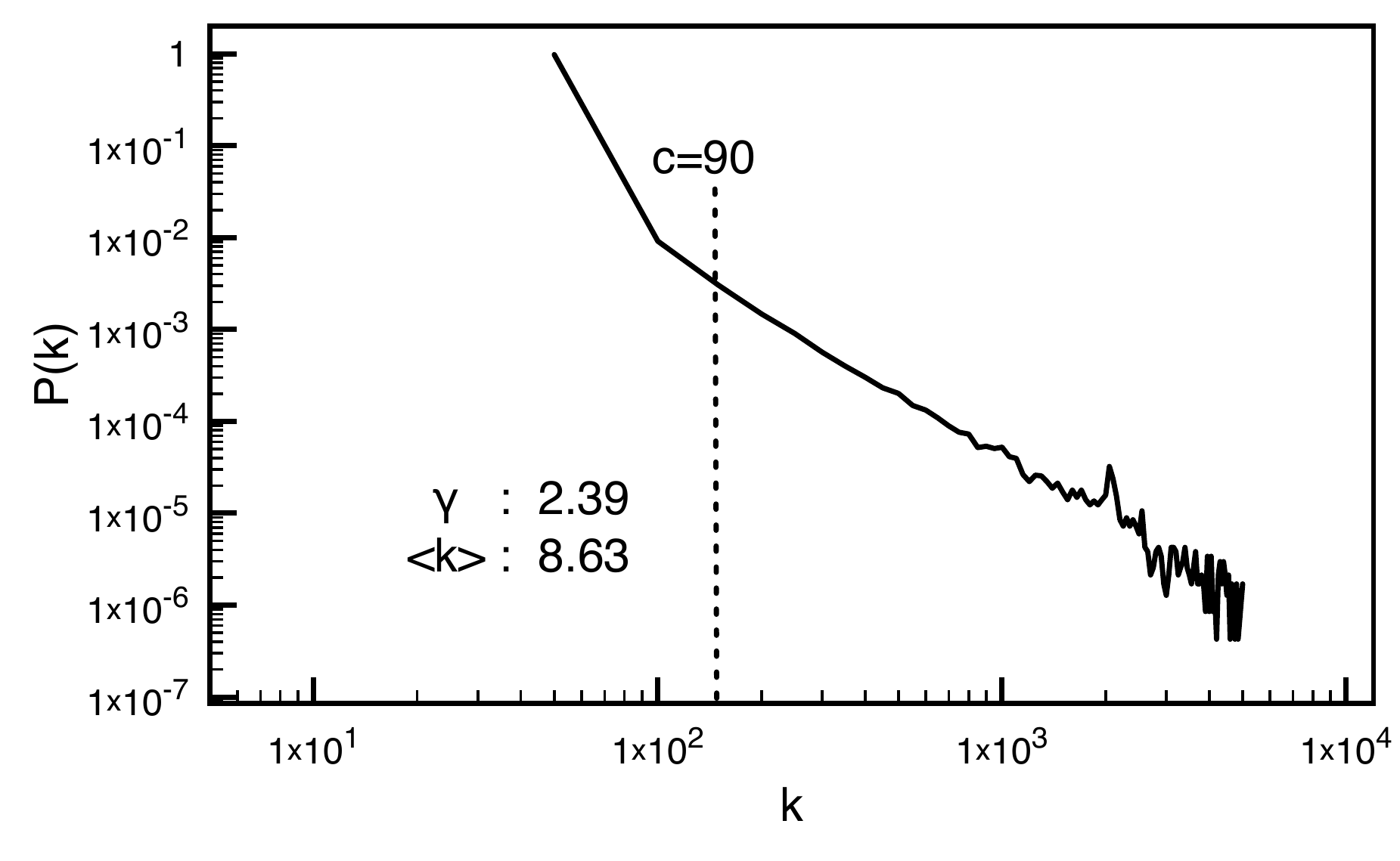}
		\caption{Twitter Follower Network \cite{Cha2010}}
		\label{fig:twitterfollower}
	\end{subfigure}
	~ 
	\begin{subfigure}[t]{0.45\textwidth}
		\centering
		\includegraphics[scale=0.4]{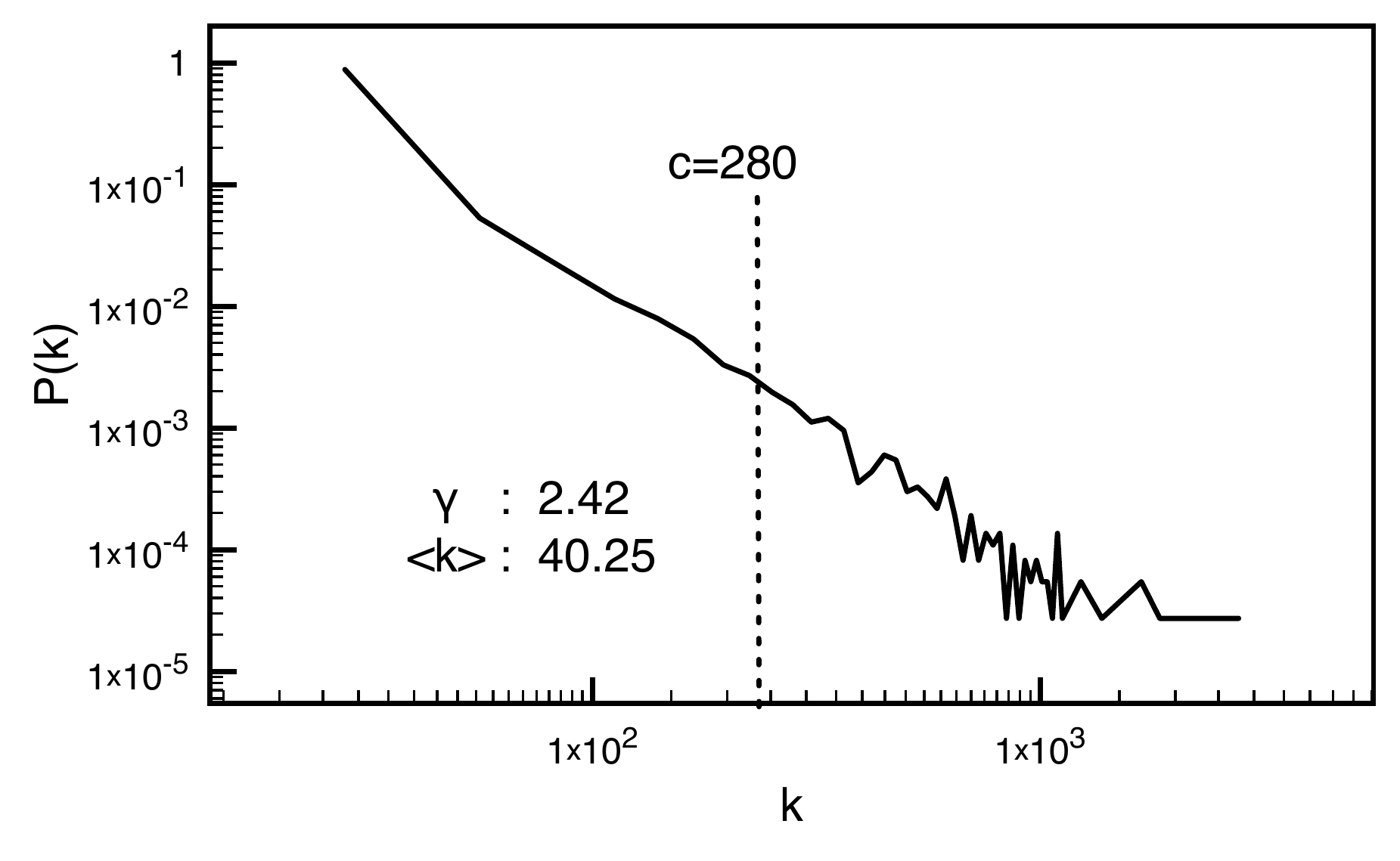}
		\caption{Enron Email Communication Network \cite{Leskovec2011}}
		\label{fig:enron}
	\end{subfigure}
	~ 
	\begin{subfigure}[t]{0.45\textwidth}
		\centering
		\includegraphics[scale=0.4]{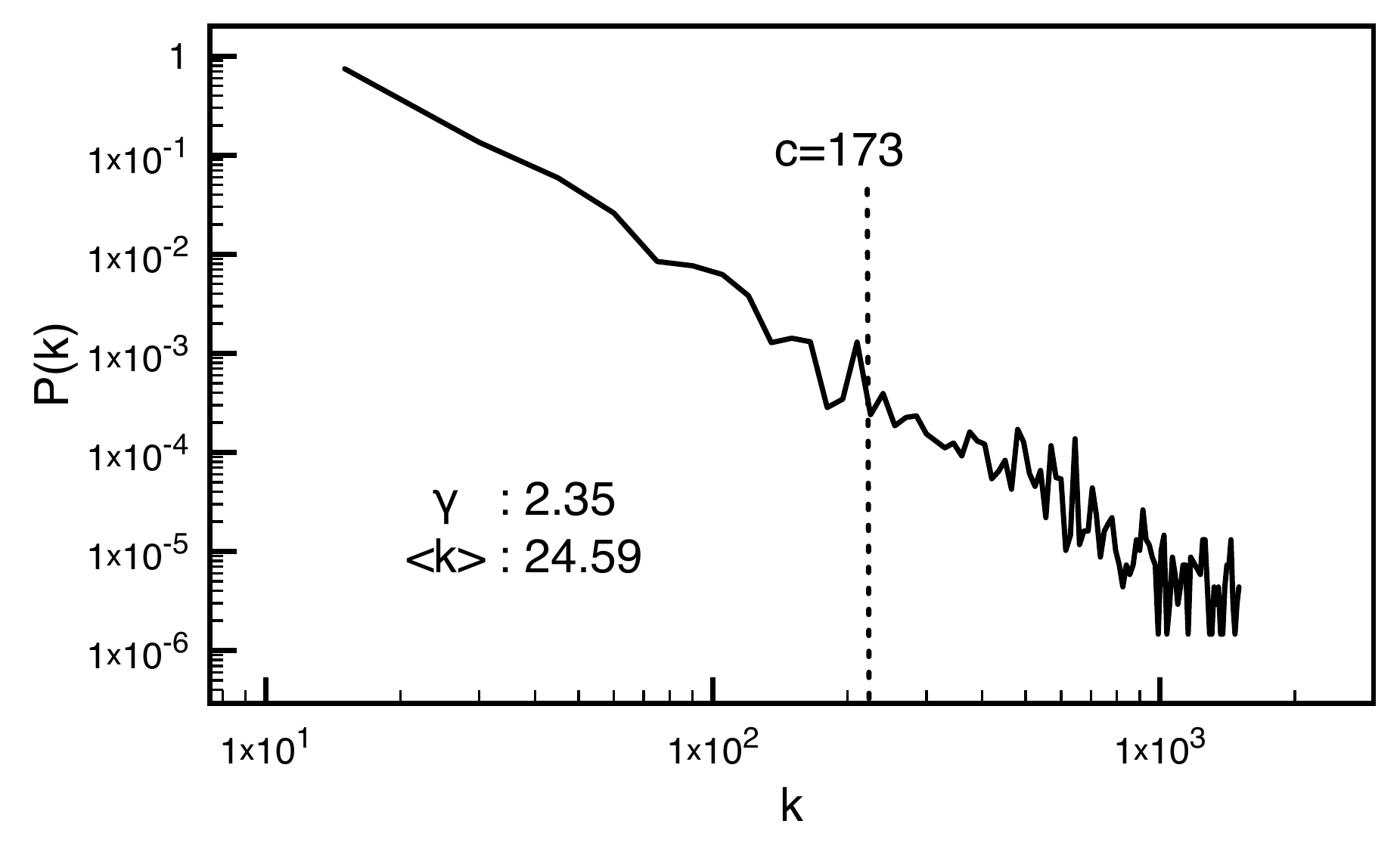}
		\caption{Berkley Stanford Web Interconnection Network \cite{Leskovec2011}}
		\label{fig:berkstan}
	\end{subfigure}
	\caption{Degree Distributions from Social Networking and Web Networks on a Logarithmic Scale}
	\label{fig:social-degree}
	\vspace{-1.5em}
\end{figure*}

\begin{figure*}[t]
	\centering
	\begin{subfigure}[t]{0.45\textwidth}
		\centering
		\includegraphics[scale=0.4]{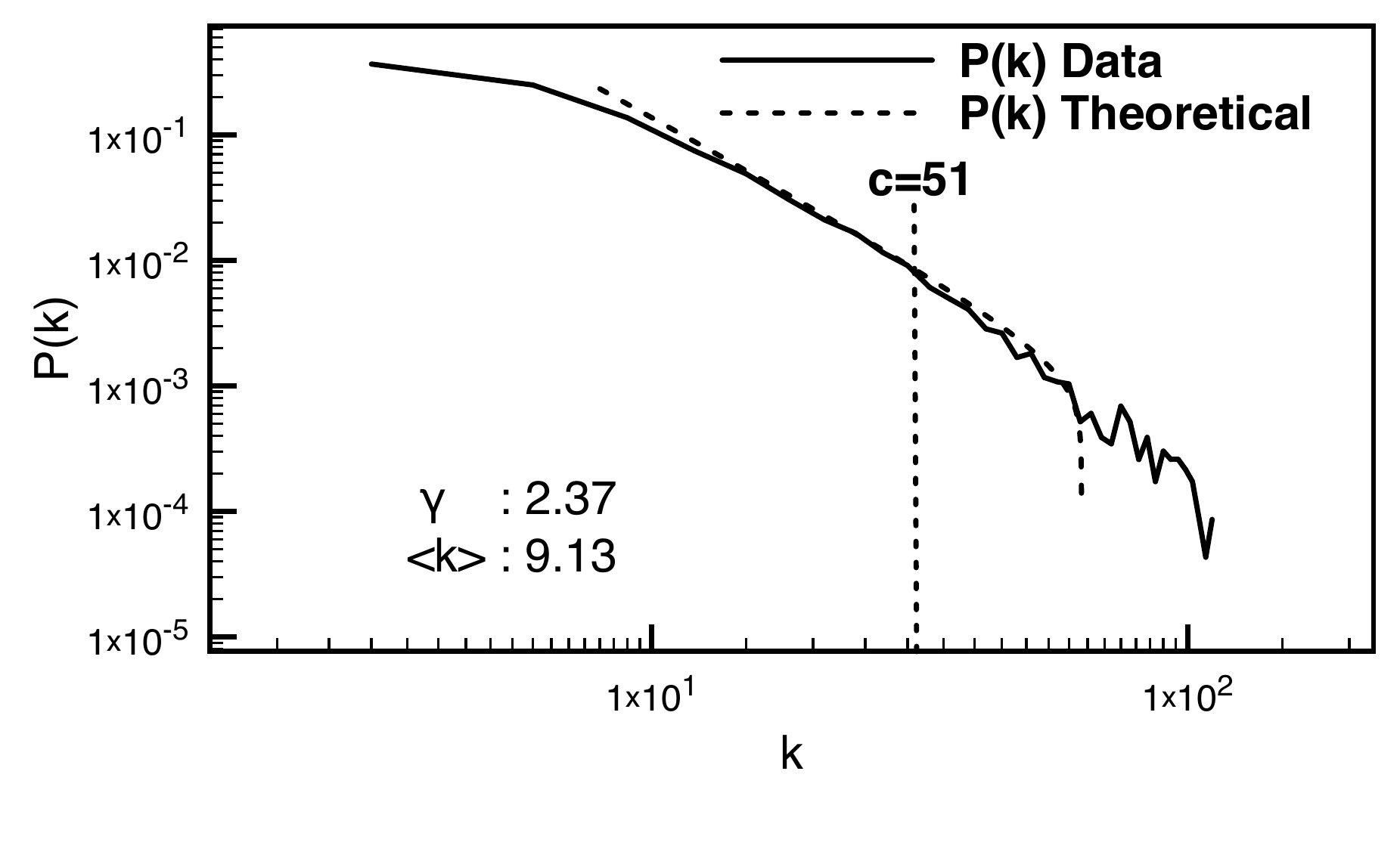}
		\caption{Arxiv Condensed Matter Citation Network, Theoretical and Experimental \cite{Leskovec2007}}
		\label{fig:arxiv-condmatt}
	\end{subfigure}
	~ 
	\begin{subfigure}[t]{0.45\textwidth}
		\centering
		\includegraphics[scale=0.4]{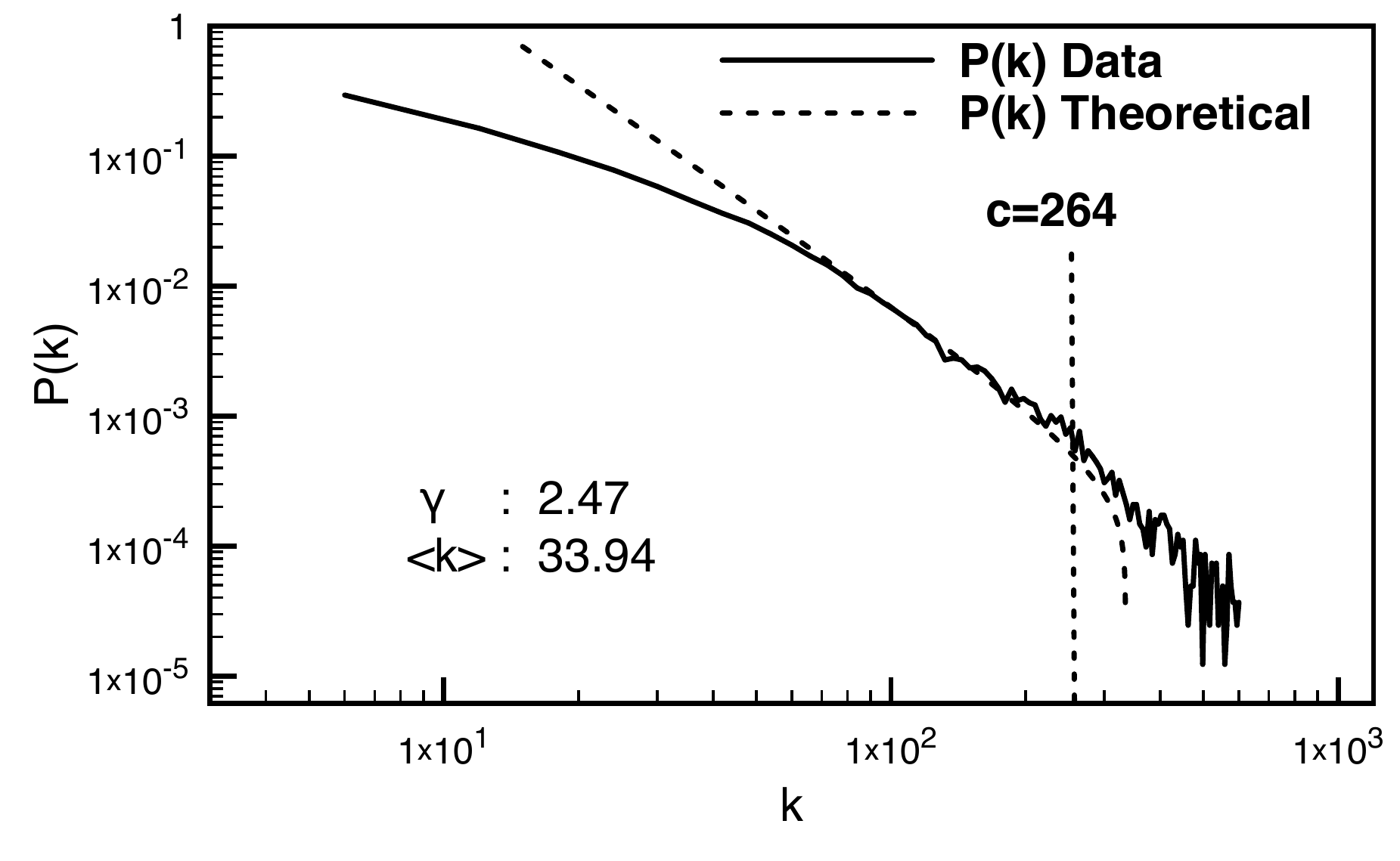}
		\caption{Patent Citation Network,  Theoretical and Experimental \cite{Leskovec2007}}
		\label{fig:patent}
	\end{subfigure}
	~
	\begin{subfigure}[t]{0.45\textwidth}
		\centering
		\includegraphics[scale=0.4]{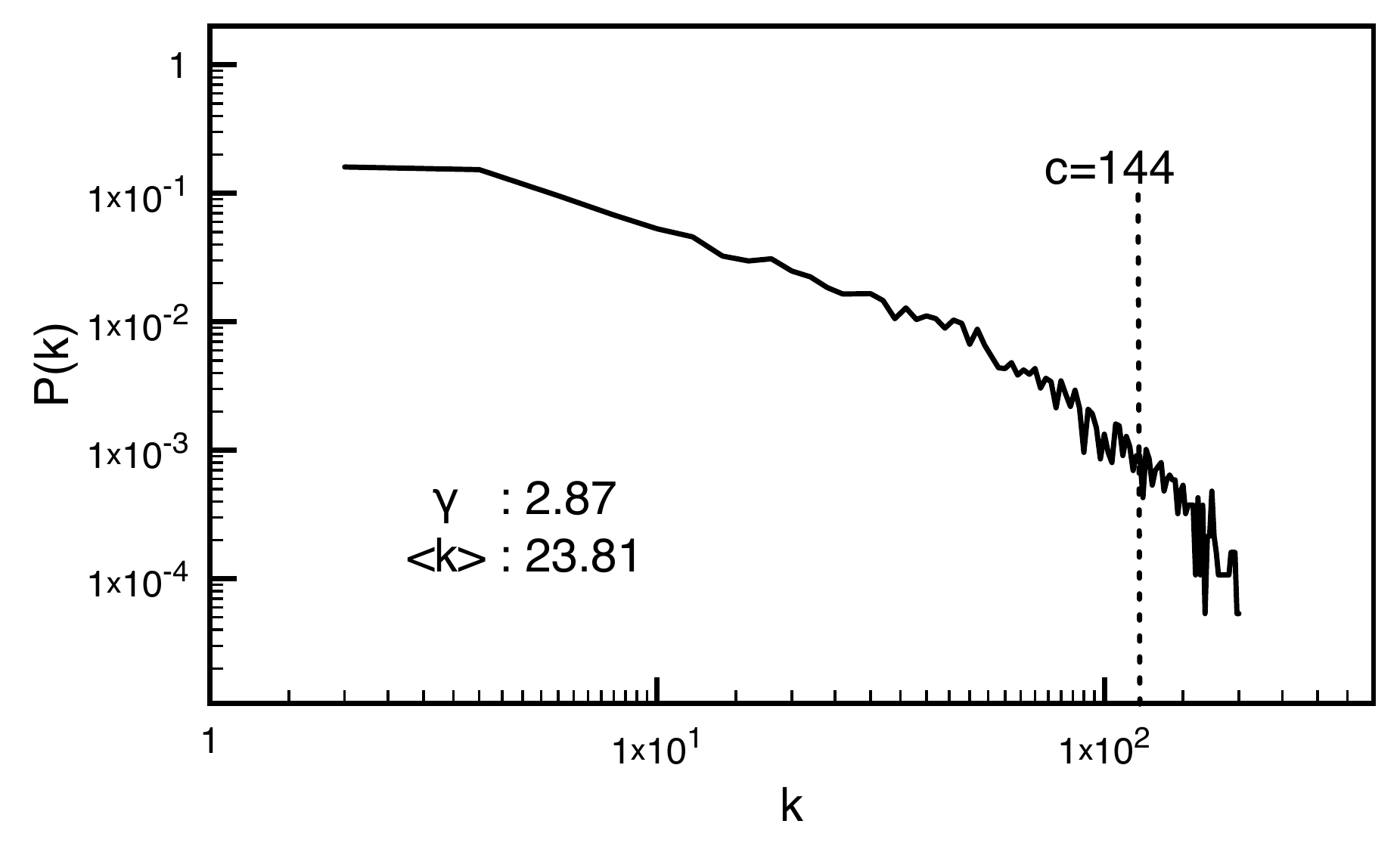}
		\caption{Arxiv Astro-Physics Citation Network \cite{Leskovec2007}}
		\label{fig:axiv-astro}
	\end{subfigure}%
	~ 
	\begin{subfigure}[t]{0.45\textwidth}
		\centering
		\includegraphics[scale=0.4]{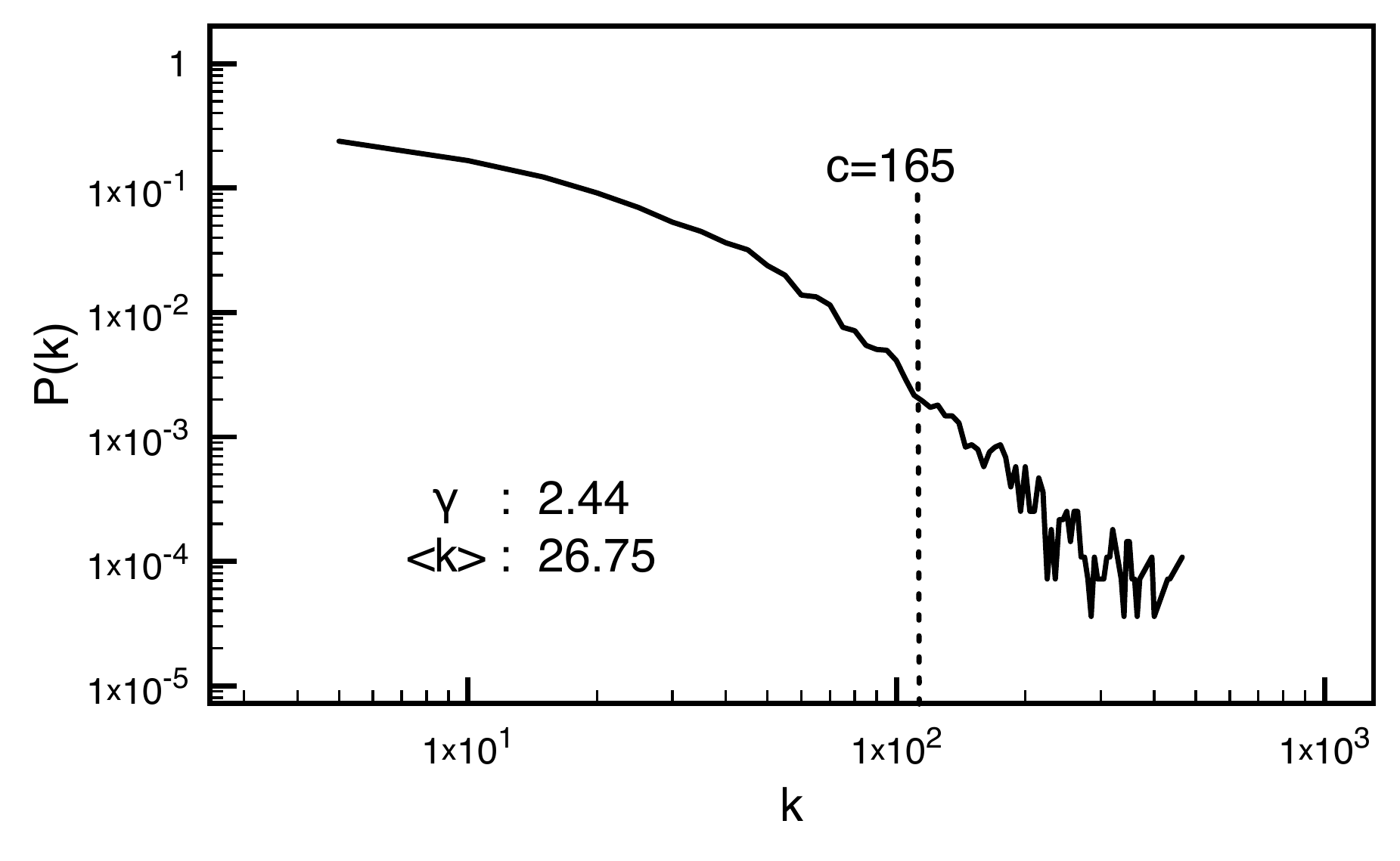}
		\caption{Arxiv High Energy Physics Citation Network \cite{Leskovec2007}}
		\label{fig:arxiv-hepth-cit}
	\end{subfigure}
	~ 
	\begin{subfigure}[t]{0.45\textwidth}
		\centering
		\includegraphics[scale=0.4]{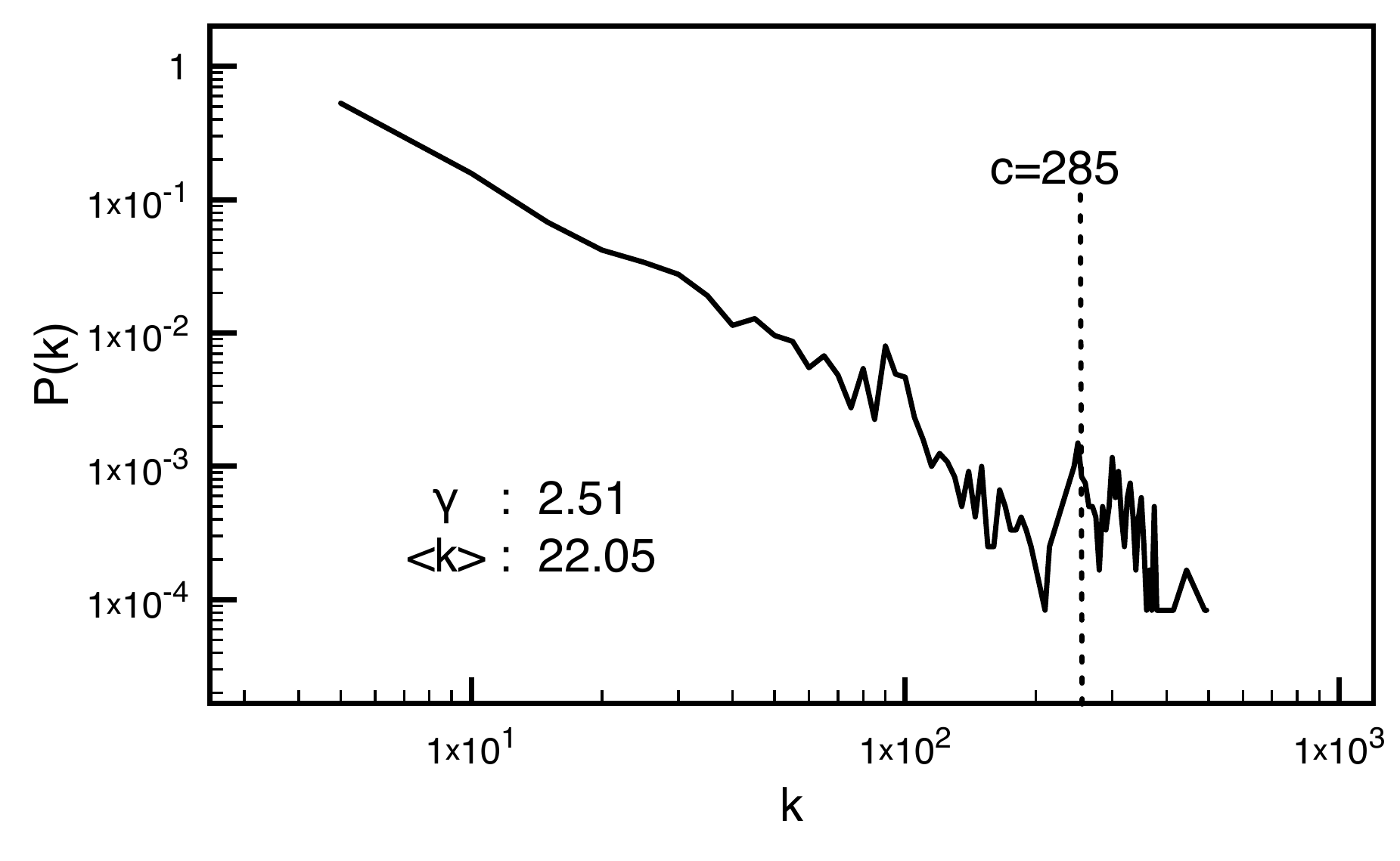}
		\caption{Arxiv High Energy Physics Collaboration Network \cite{Leskovec2007}}
		\label{fig:arxiv-hepth-collab}
	\end{subfigure}
	\caption{Degree Distributions from Collaboration and Citation Networks on a Logarithmic Scale}
	\label{fig:citation-degree}
	\vspace{-1.5em}
\end{figure*}

\begin{figure*}[t]
	\centering
	\begin{subfigure}[t]{0.45\textwidth}
		\centering
		\includegraphics[scale=0.4]{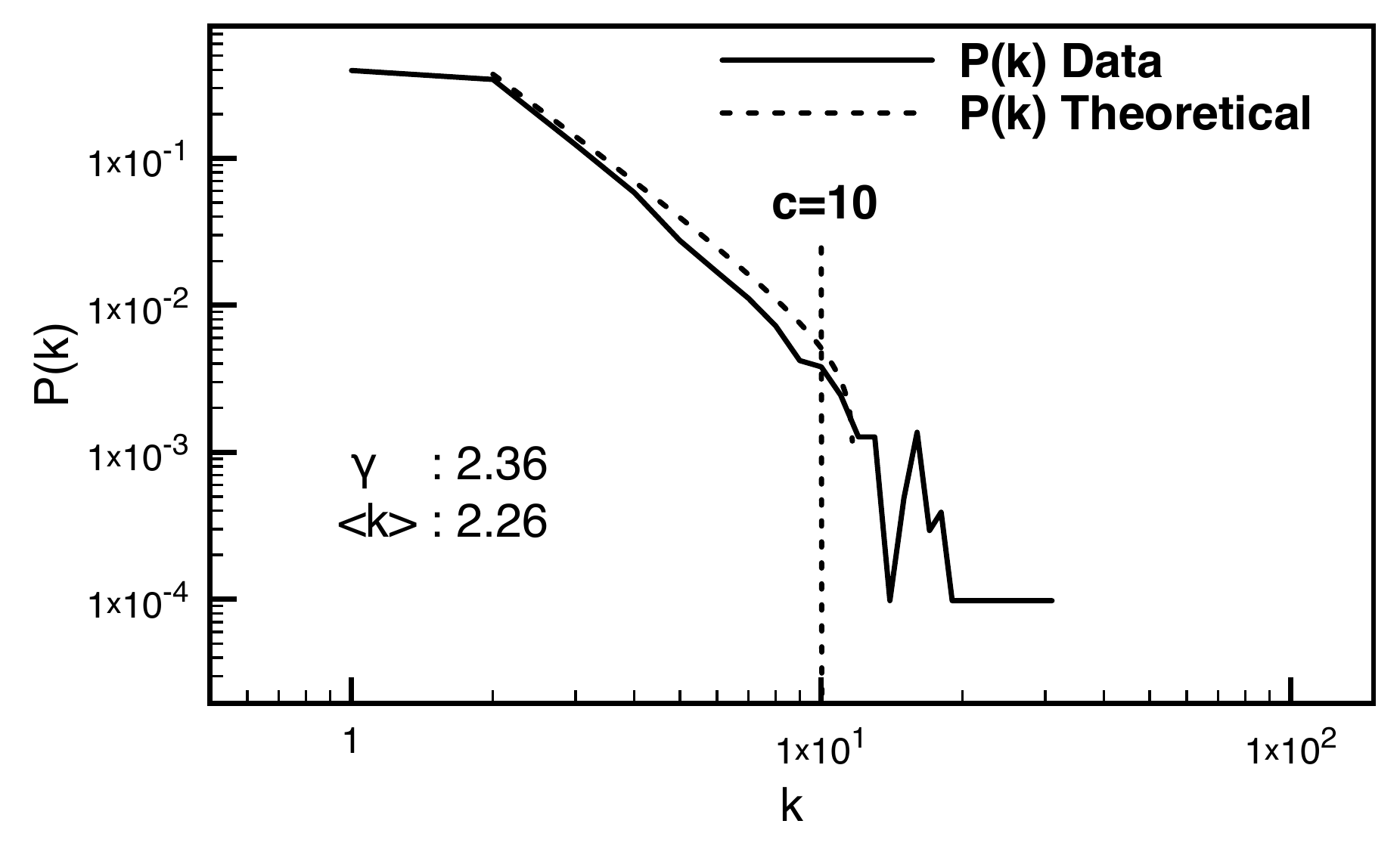}
		\caption{Internet Topology Zoo Network, Theoretical and Experimental \cite{Knight2011a}}
		\label{fig:itzoo}
	\end{subfigure}
	~ 
	\begin{subfigure}[t]{0.45\textwidth}
		\centering
		\includegraphics[scale=0.4]{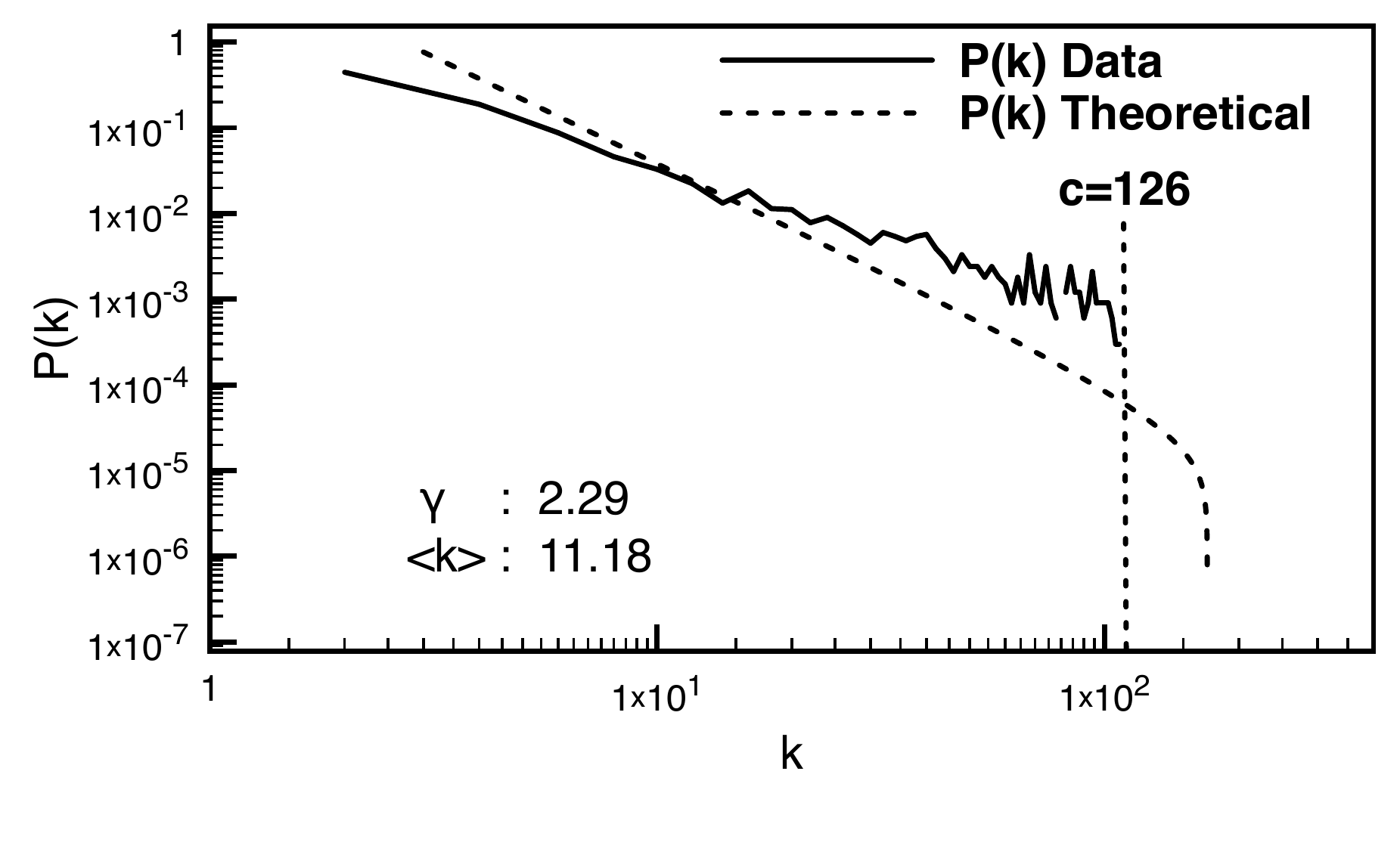}
		\caption{Airport Flight Interconnection Network, Theoretical and Experimental \cite{Patokallio2016}}
		\label{fig:airports}
	\end{subfigure}
	~ 
	\begin{subfigure}[t]{0.45\textwidth}
		\centering
		\includegraphics[scale=0.4]{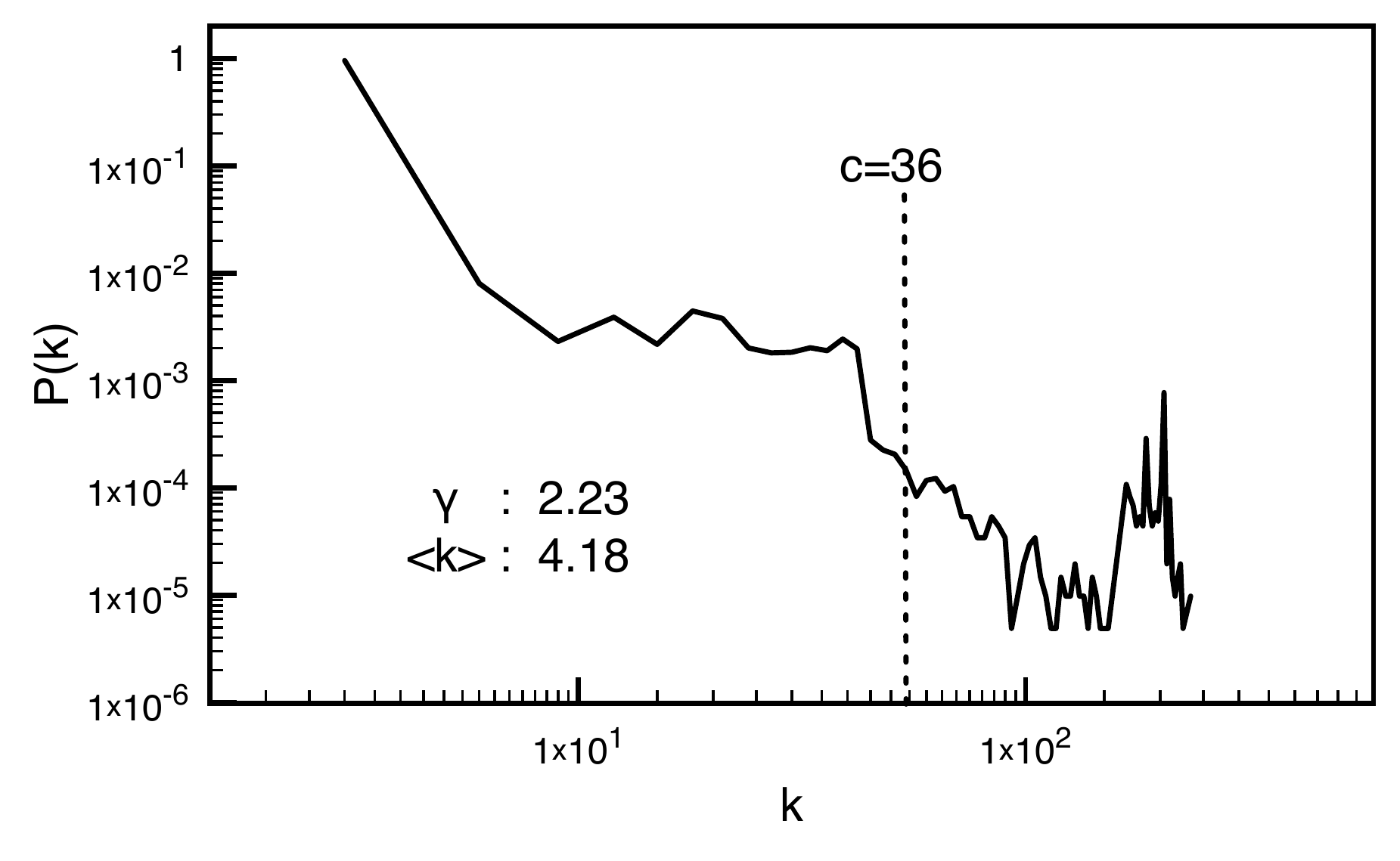}
		\caption{Web Provider Datacenter Network \cite{Tee2016b}}
		\label{fig:web}
	\end{subfigure}
	~ 
	\begin{subfigure}[t]{0.45\textwidth}
		\centering
		\includegraphics[scale=0.4]{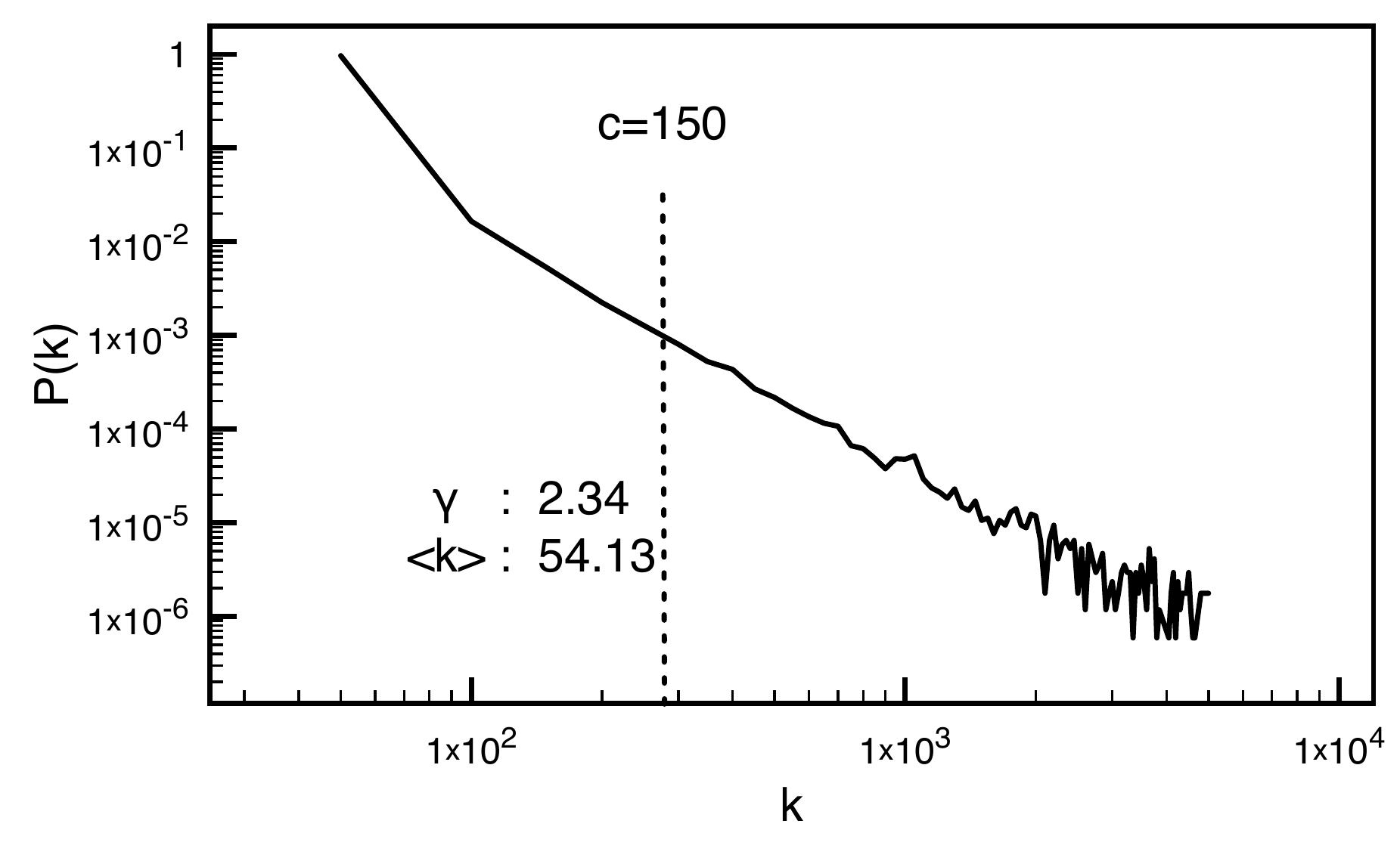}
		\caption{Internet Autonomous Systems Network  \cite{Leskovec2007}}
		\label{fig:asskitter}
	\end{subfigure}
	\caption{Degree Distributions from Infrastructure and Communications Networks on a Logarithmic Scale}
	\label{fig:infra-degree}
	\vspace{-1.5em}
\end{figure*}

We also present the analysis both as a collection of log/log distribution graphs in Figures \ref{fig:social-degree}, \ref{fig:citation-degree} and \ref{fig:infra-degree} and also summarize the key prediction of $\gamma$ against the standard value of $3.0$ from preferential attachment in Table \ref{tab:gammatab}. In the log/log plots we overlay the value of $c$ at $99^{th}$ percentile, the average value of $\gamma$ to this constraint and the expected value of the node degree $\langle k \rangle$. In each of Figures \ref{fig:social-degree}, \ref{fig:citation-degree} and \ref{fig:infra-degree}, we also overlay the theoretical prediction for the distribution $P(k)$ obtained by substituting the values of $\gamma$ from Table \ref{tab:gammatab} into Equation (\ref{eqn:main_result}). The agreement between the predicted values of $\gamma$ and the measured ones for our datasets is evident from these combined theoretical and experimental plots, at least for portions of the distribution. A consequence of the selection of $c$ at the $99^{th}$ percentile is that our theoretical curve displays a cut off earlier than the experimental data, which is to be expected.

The striking feature of many of the degree distributions is the absence of strict linearity, contrary to the predictions of the standard scale free model, and also the marked increase in $\gamma$ at high values of $k$, a key prediction of our constrained model and a necessary precursor to a hard constraint in the value of $k$. In the social network data we analyzed this is best illustrated in Figures \ref{fig:pokec}, \ref{fig:facebook} and \ref{fig:twittercircles}. Similar behavior is also present in the citation network (perhaps the best example being Figure \ref{fig:arxiv-hepth-cit}), and again in the infrastructure graphs, particularly the Internet Topology Zoo (Figure \ref{fig:itzoo}). It is interesting to speculate what the nature of the constraint is in the social networks, but this is perhaps explained by the effective limitations, no matter how small, on the amount of time people can feasibly spend on social networking platforms. Indeed in almost every conceivable network a constraint is a natural feature. Whether the node in the graph is a physical device, and individual engaged in an activity such as writing papers, or web site hyperlinks, there is a limitation to the connections a node can have. In some cases these are hard design limits such as ports on a network switch, in others it is simply the capacity of a human being, with a fixed lifespan, to blog, interact, star in a movie or engage in any other social activity. In every case our experimental data bears this out.

In the following Section \ref{sec:dynamics} we point out how the two models may well be related to a fundamental dynamical principle that arises from thermodynamic considerations of network evolution. Critically this analysis derives the form of preferential attachment presented as an axiom in the scale free model.

%
%

%
%

%
%

\section{Dynamical Evolution of Scale Freedom}
\label{sec:dynamics}
In our treatment thus far we have followed the continuum model of Barab{\'a}si-Albert with the addition of a constraint-based factor to the attachment probability. 
However, we can attack the problem from a more fundamental viewpoint. 
Essentially, we argue that the evolution of a graph satisfies the criteria for a treatment based upon considerations of entropy from a statistical mechanics perspective, in accordance with the $2^{nd}$ law of thermodynamics. In any isolated physical system the entropy of the system will tend to a maximum unless energy is input  to prevent that. For a classic treatment see \cite{Schrodinger1989}. 
In natural processes this tendency to increase entropy can be modeled as a macroscopic force on the system. This entropic force is responsible for both the elasticity of certain polymers and the biological process of osmosis. Indeed if thermodynamic temperature is written as $\textbf{T}$ and entropy $\textbf{S}$, one can state the entropic force $F$ acting on a body when a process changes entropy as follows:

\begin{equation}\label{eqn:entropic_force}
	F=\textbf{T} \Delta \textbf{S} \mbox{ .}
\end{equation}

 To begin our treatment of graph evolution from fundamental thermodynamic principles, it suffices to pose the problem in an appropriate manner. Consider an existing graph of $m_{0}$ nodes and $e_{0}$ edges in thermal equilibrium with an infinite supply of unattached nodes, each capable of connecting to $m$ nodes in the event that it comes into contact with the existing graph. 
At every time-step we imagine that such an interaction occurs and the new node connects to $m$ others. Our problem is to identify the probability of attachment for a node according to its degree $k$, and thus derive the degree distribution. More strictly, it is necessary to consider an ensemble of all possible graph configurations, \emph{at every time step}, to enable statistical treatment of this process. 
This requirement to consider an ensemble of configurations is at first sight an added complication, but in fact is critical in permitting the analysis of the model. Whenever we consider a randomly selected node, for example in equation (\ref{eqn:dynamic_pi}), it is important to recognize that we must average any interaction with the remaining graph over \emph{all possible graphs} that can be constructed from the subgraph obtained by removing the randomly selected node and all edges connected to it. This ensemble average is further constrained by the total number of vertices and edges being unchanged after the removal of the random node.
This requirement to average over all possible graph configurations at each time step justifies the approximation we make to calculate, for example, the average clustering coefficient.

The probability of attachment to a random node must statistically and universally seek to maximize total entropy. Our model proposes that the probability of this random node acquiring new links is a result of the relative strength of the entropic force of attachment to the randomly chosen node versus any other node in the graph. Those nodes which exert the highest entropic force relative to the rest of the nodes in the network will gain the most links, and we write this mathematically as:

    \begin{equation}\label{eqn:dynamic_pi}
        \Pi_{i} = \frac{F(v_{i})} { \sum_{j \neq i} F(v_{j}) }
    \end{equation}
    where $F(v_{i})$ is the entropic force of attraction to node $i$.

This expression governs the individual interaction that our randomly selected node has with a particular graph configuration,  analogous to the elastic collision equations used to formulate the statistical treatment of ideal gases. In a similar way we cannot easily analytically formulate the dynamical equations of the graph from this equation as they are very large, and so to derive the degree evolution equations from this formulation we utilize statistical ensemble arguments. Considering all possible configurations of the graph $G(V(t), E(t))$ at a fixed time $t$, the denominator of equation (\ref{eqn:dynamic_pi})  is computed as an expectation value of the relative force of attaching to any other node, across all possible graphs at time $t$ in the ensemble that our random node could be connected to. At a given time $t$ in the evolution of the graph the numbers of vertices $|V(t)|$ and edges $|E(t)|$ are constant, but we do have to consider all possible graph configurations of that number of vertices and edges. This will ultimately change the average of the change in entropy that the node could make on connecting to any other node in the graph other than our randomly selected node $v_{i}$. In this way we collapse the denominator to the expected value of this entropy change, averaged across all possible connection points in all possible members of the ensemble. We write this as $\textbf{T} \times |V| \times \mathbb{E} ( \Delta \textbf{S} )$. As the graph becomes larger, we make the assumption that  the value of $|V| \times \mathbb{E} ( \Delta \textbf{S} )$ is effectively constant, and factor this out. We base this assumption on the fact that most real world networks do indeed demonstrate some form of steep drop in the distribution of node degrees, so that the vast majority of nodes posses low degree (an important claim of \cite{Watts1998} and \cite{Albert2002}). It seems reasonable to assume that with such a restricted degree sequence most nodes will contribute a similar amount to the change in entropy, and this expected value will stabilize to a constant. More complex analysis could admit a time varying value of this constant, as strictly both $V$ and $\mathbb{E} ( \Delta \textbf{S} )$ may have complex time dependence, but for simplicity we assume:
    
    \begin{equation*}
        \epsilon=\frac{1}{|V| \times \mathbb{E} ( \Delta \textbf{S} ) } \mbox{ .}
    \end{equation*}

    With this assumption equation (\ref{eqn:dynamic_pi}) simplifies and $\textbf{T}$ factors out to yield

    \begin{equation}
        \Pi_{i} = \epsilon \Delta \textbf{S}_{i} \mbox{ .}
    \end{equation}

In general $\textbf{S}_{i}$ is a function of potentially many variables $x_{i}$, but certainly depends upon $k_{i}$ and time $t$. We can calculate $\Delta \textbf{S}_{i}$ as a total differential, $\Delta \textbf{S}_{i}(x_{j}) = \sum\limits_{x_{j}} \frac{ \partial \textbf{S}_{i} }{ \partial x_{j} } \Delta x_{j}$, but we can assume for simplicity that $t$ is fixed and the dependence is purely upon $k_{i}$. In this case $\Delta \textbf{S}_{i} = \frac{d \textbf{S}_{i} }{ d k_{i} } \times \Delta k_{i}$, with,  for a single time step, $\Delta k_{i} = 2m$. This gives us our expression for attachment probability:

\begin{equation}\label{eqn:entropy_pref}
    \Pi_{i} = \epsilon 2m \frac{d \textbf{S}_{i}(k_{i}) }{ d k_{i} } \mbox{ .}
\end{equation}

To make use of equation (\ref{eqn:entropy_pref}) we require an expression for the entropy of a node in the graph. The subject of the entropy of a graph has a long history, originating in the work of K{\"o}rner on the informational entropy of signals described in \cite{Korner1986} and \cite{Simonyi1995}. Many approaches to calculating the entropy of a graph have been proposed, including the use of the eigenvalues of the adjacency matrix (see \cite{Passerini2008}, and ensembles of networks with similar degree sequences (proposed in \cite{Bianconi2008}).  Unfortunately these concepts relate to the global value of entropy for a graph, and do not have utility when calculating the change in entropy as a new node connects.

A series of papers by Dehmer (\cite{Dehmer2008},\cite{Dehmer2011}) formalized the concept of the individual entropy of a node. In recent work \cite{Tee2016b} we built upon this formulation to define a local vertex measure (referred to in \cite{Tee2016b} as $NVE'$, and equivalent to our definition of $\textbf{S}_{i}$ here) in terms of its relative degree as:

\begin{equation}\label{eqn:vertex_ent}
    \textbf{S}_{i}(k_{i},t)=\frac{1}{C_{i}^{1}} \times \frac{k_{i}}{2|E(t)|} \log \frac {2|E(t)|}{k_{i}} \mbox{ ,}
\end{equation}

where $C_{i}^{1}$ represents a modified clustering coefficient of the 1-hop neighborhood of the node $v_{i}$. Contrary to the more common point-deleted neighborhood clustering coefficient, $C_{i}^{1}$ preserves the node in the calculation to measure similarity to the local perfect graph $K_{n}$ of order $n=k_{i} +1$.  For convenience we give an explicit definition of the 1-hop neighborhood $N_i^1$:

    \begin{equation*}
        N_{i}^{1}= \{ v \in V \ | \  d(v_{i},v) \leq 1 \} \cup \{ v_{i} \} \mbox{ ,}
    \end{equation*}
    and the related `1-edges' $E_{i}^{1}$ as
    \begin{equation*}
      E_{i}^{1}=\{ e_{jk} \in E \ | \ v_{j} \in N_{i}^{1} \mbox{ and  } v_{k} \in N_{i}^{1} \} \mbox{ .}
    \end{equation*}

    We can then define the modified clustering coefficient to be

\begin{equation}
        C_{i}^{1}=\frac{ 2 |E_{i}^{1}|}{k_{i}(k_{i}+1)} \mbox{ .}
\end{equation}

At this point we can make use of the fact that we must consider all possible intermediate graph configurations to assume effective uniformity in the graph to calculate $|E_{i}^{1}|$, and assert that for a given node, $|E_{i}^{1}| = \frac{k_{i}+1}{|V|} \times |E(t)|$. This then yields for the clustering coefficient the following expression:

\begin{equation}
	C_{i}^{1}=\frac{ 2|E(t)| }{ k_{i}|V(t)|} \mbox{ .}
\end{equation}

Given that at every time-step we add one node to the graph, connecting to $m$ other nodes we can write $|V|=m_{0}+t$, and $|E|=e_{0}+mt$. In general as the model evolves, $t \gg m_{0}$ and similarly, $mt \gg e_{0}$, these simplify to $|V|=t$ and $|E|=mt$. Substituting back in we obtain the following equation for vertex entropy at $v_{i}$ at time $t$ as:

\begin{equation}\label{eqn:vertexent_time}
	\textbf{S}_{i}(k_{i},t)=\frac{ k_{i}^{2} }{ 4m^2 t } \log \bigg ( \frac{ 2mt }{k_{i}} \bigg ) \mbox{ .}
\end{equation}

In the analysis undertaken by Tee \emph{et al} in \cite{Tee2016b,Tee2017}, this quantity was identified as sharing some of the properties of the structural entropy of the graph when summed across all vertices.  In  particular, the extremal behavior of the summed vertex entropy was proven to be minimized by the perfect graph of order $n$, $K_{n}$, and maximized by the star graph of order $n$, $S_{n}$, for simply connected undirected graphs. From the perspective of dynamical evolution of networks, this is consistent with the approach in our analysis. The perfect graph $K_{n}$ will tend towards a more node level disordered graph such as $S_{n}$ as addition of nodes selects targets such as to increase the value of $\textbf{S}_{i}$ in Equation (\ref{eqn:vertexent_time}). From a purely statistical mechanics perspective one can consider each connected graph on $n$ nodes and $|E|$ edges as representing a micro-state. The perfect graph is achievable in precisely one unique configuration if edges are indistinguishable, whereas other configurations, $S_{n}$ for example, can be achieved by selecting any one of the nodes as the hub vertex. In this way the result that increases in entropy tends to destroy cliques and regular ordered graphs is consistent. From this perspective we would expect dynamic processes to favor the attachment to nodes where the increase in $\textbf{S}_{i}$ is greatest. From here it is straightforward to follow through the continuum analysis as described in \cite{Albert2002}. For the time evolution of $k$ the following equation, is obtained:

\begin{equation}\label{eqn:entropic_attach}
	\frac{dk_{i}}{dt} = 2m\Pi_{i} = -\epsilon \frac{k_{i}}{t} \Bigg \{ \frac{1}{2} +  \log \bigg ( \frac{k_{i}}{2mt} \bigg ) \Bigg \} \mbox{ .}
\end{equation}
Although at first sight this nonlinear ODE appears intractable, in fact
an analytic solution is available. Making the change of variables
$y=\log k$ and $x=\log t$, so that
$\frac{dy}{dx} = \frac{t}{k} \frac{dk}{dt}$, we see
that~\eqref{eqn:entropic_attach} becomes
\begin{equation}
\frac{dy}{dx} = -\epsilon \left[ \frac{1}{2} + y - \log(2m) -x
\right] \nonumber
\end{equation}
This is now a linear ODE which can be solved by standard methods.
Applying the initial condition $k_i(t_i)=m$ the solution is found to be
most conveniently expressed in the form
\begin{equation}
\begin{split}
\log k_i(t) = \log (2mt) - \frac{1}{2} - \frac{1}{\epsilon} + \\
\left[ 
\frac{1}{2} + \frac{1}{\epsilon} - \log(2t_i) \right] 
\left( \frac{t_i}{t} \right)^\epsilon 
\end{split}
\label{eqn:ent}
\end{equation}
For values of $\epsilon<1$ the behavior of $k_i(t)$
is similar to the Barabasi--Albert model: degrees increase monotonically
but at an ever decreasing rate. An analytic form for the degree
distribution, analogous to~\eqref{eqn:ba} does not seem straightforward
to derive.

\begin{figure*}[t]
	\centering
	\begin{subfigure}[t]{0.45\textwidth}
		\centering
		\includegraphics[scale=0.5]{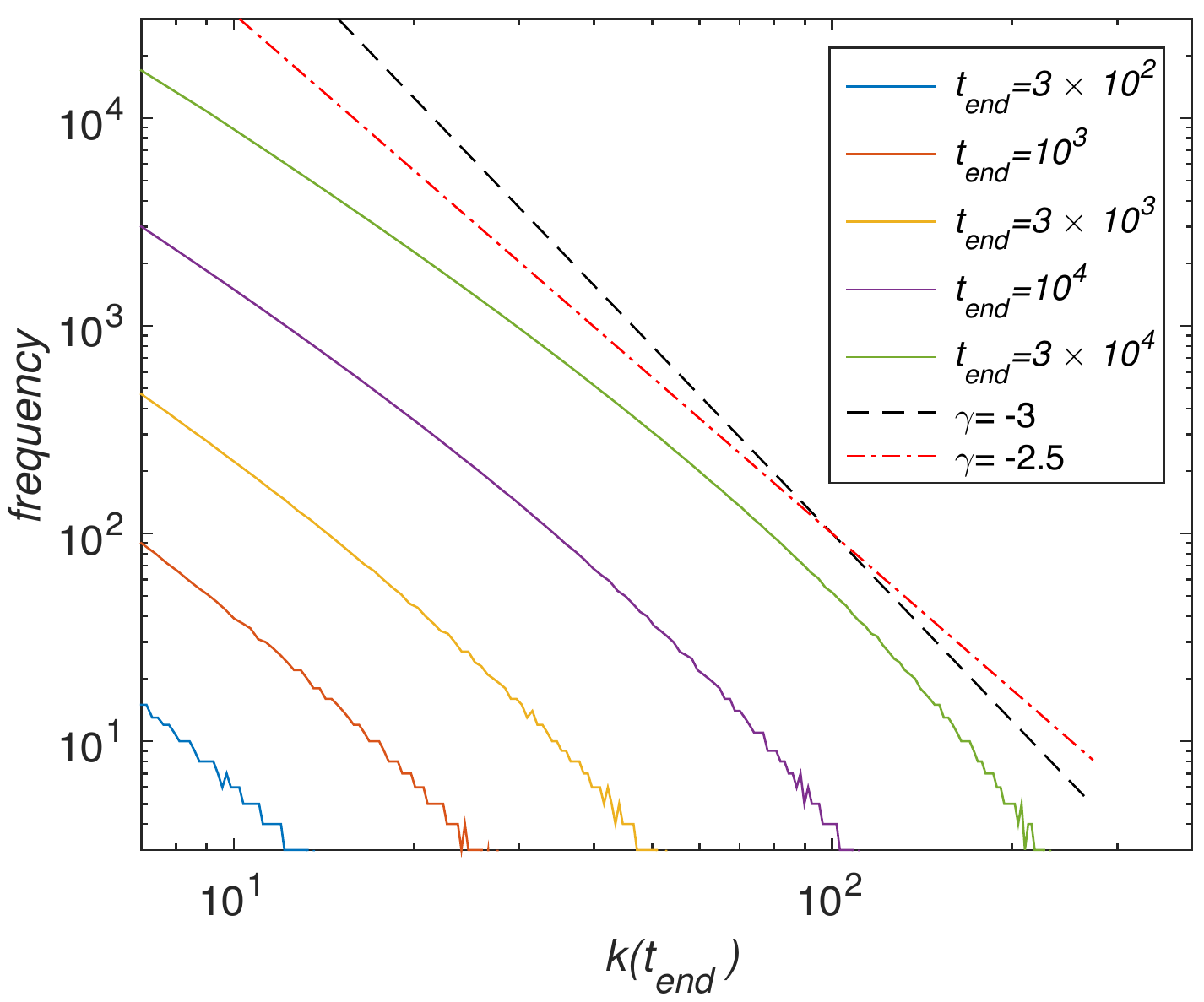}
		
\centerline{(a)}
	\end{subfigure}%
	~ 
	\begin{subfigure}[t]{0.45\textwidth}
		\centering
		\includegraphics[scale=0.48]{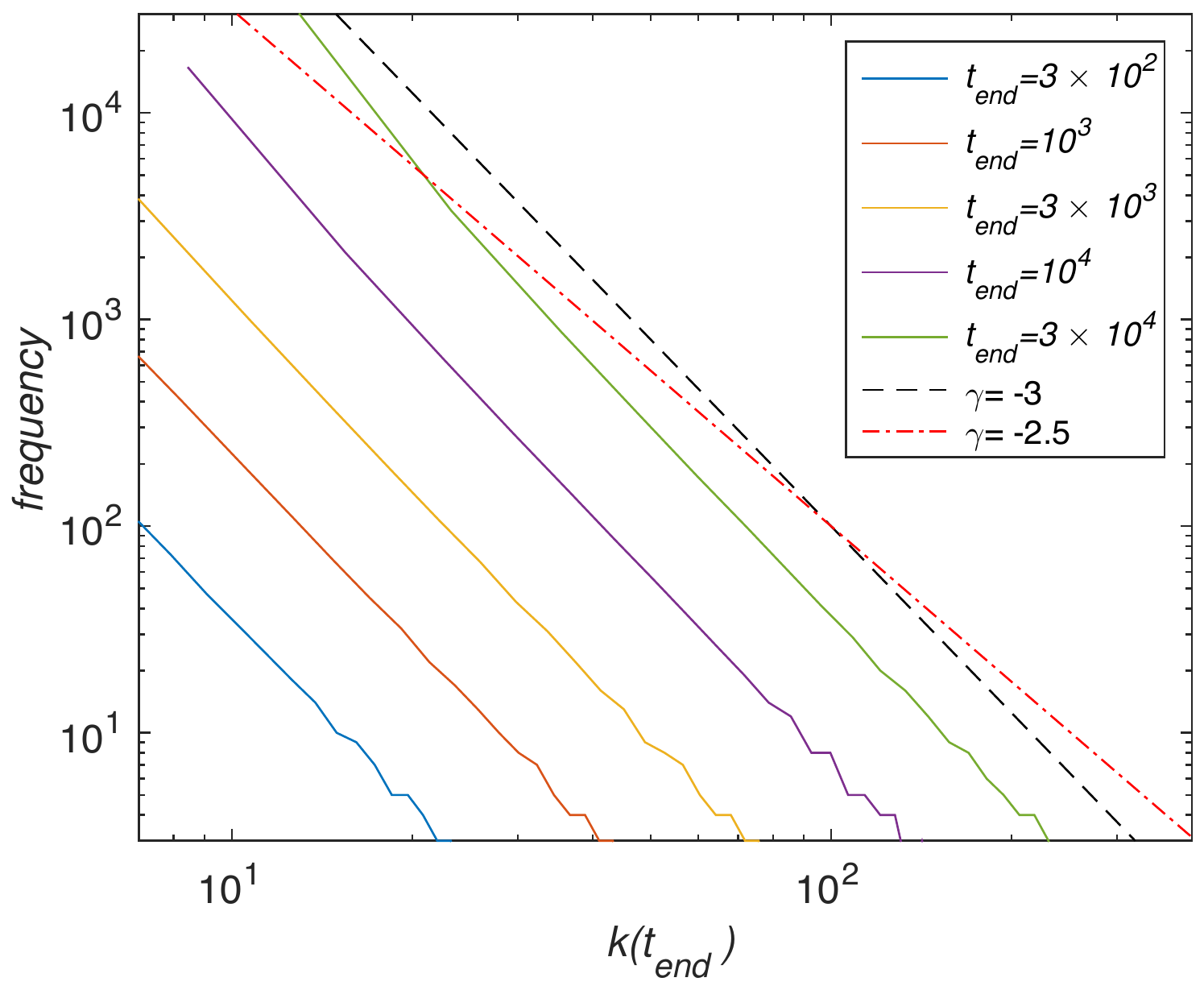}
\centerline{(b)}
	\end{subfigure}
	\caption{Degree distributions for growing networks
	at fixed end times $t_{end}=3\times 10^2, 10^3, 3\times 10^3, 10^4$,
	and $3 \times 10^4$. (a) Entropy-based model. (b) Barabasi--Albert
	model. In both cases a new node is
	introduced every $\Delta t=0.5$ and the node degrees evolve according
	to~\eqref{eqn:ent} or~\eqref{eqn:ba}, in (a) and (b) respectively.
	Parameter values: $m=5$, $\epsilon=0.1$. For illustration we have plotted the power law distribution lines at $\gamma=-3$ and $-2.5$.}
	\label{fig:comp}
	\vspace{-1.5em}
\end{figure*}

Figure~\ref{fig:comp} compares numerically computed
degree distributions from the model~\eqref{eqn:ent}
(shown in figure~\ref{fig:comp}a) and the Barabasi--Albert
model, shown in figure~\ref{fig:comp}b.
In each case a new node was added to the network every 0.5 time units,
setting $m=5$ and growing the degrees of existing nodes according
to~\eqref{eqn:ent} or~\eqref{eqn:ba} respectively. Degree distributions
are plotted for fixed end times $t_{end}$, taking 
the values $3\times 10^2$, $10^3$, $3\times 10^3$, $10^4$, and
$3 \times 10^4$. The degree distributions for the entropy-based
model do not clearly follow any power law behaviour, at least
in the regime explored here, while the Barabasi--Albert model
quickly assumes a form very close to a power-law degree distribution with exponent $\gamma=3$ as we expect.

While any systematic analysis of~\eqref{eqn:ent} seems difficult,
for large enough networks we might expect that this model is comparable
to the classes of sub-linear preferential attachment models
studied rigorously by Dereich \& M\"orters \cite{DM2009,DM2011}.
These authors prove that preferential attachment rules based on
concave functions of node degree will asymptotically
result in  degree distributions with exponent $\gamma=3$. This
suggests that the long time dynamics of the entropy-based model
might also show this behavior, but at intermediate times the
more complex distributions illustrated in figure~\ref{fig:comp}(a)
might well be more typical.

\section{Conclusion and Future Directions}
\label{sec:conclusion}
In Section \ref{sec:constraints} we introduced a modification to the preferential attachment model to account for the maximum connections a node may have in a network. From the mathematical analysis we were able to predict both the value of the power law exponent $\gamma$ and the presence of a hard limit on the degree distribution. In Section \ref{sec:analysis} we applied the analysis to an extensive range of social, citation and physical infrastructure graphs, and found that the constraint model's values for $\gamma$ more accurately fitted the data. In addition, the constrained model implicitly contains a hard limit in the node degree, and the data analyzed had degree distributions with far fewer nodes of extremely large $k$ than a pure power law would predict. This is an important result because the value is arrived at as a natural consequence of the presence of constraints on the maximum node degree, rather than by introducing a distribution of additional parameters such as in the fitness model. Fitness is a valuable concept, and indeed in further work it is intended to investigate the role of a top constraint in a model extended to include the concept of fitness, or indeed generalized in a similar way to the NGF models. In particular the analogy with Bose-Einstein statistical mechanics is interesting, and opens up many applications of network science in more general theoretical physics, but the method outlined in this paper captures the essential features of real degree distributions without requiring the concept of fitness.

Motivated by the interesting results when applying concepts from statistical mechanics, and the results for vertex entropy arrived at in \cite{Tee2016b}, we also set out to see if scale free models could be arrived at from pure thermodynamic principles of entropic force. In Section \ref{sec:dynamics} we were able to obtain, from first principles, an evolution equation for the degree of a random node, which although soluble analytically, presents challenges when deriving the degree distributions according to the continuum analysis. 
The Taylor series for $log(x)$ converges only for values of $x$ in the range  $0 < x \leq 2$, but as $k \leq 2mt$, and, both terms are always strictly positive, we can safely expand the log term in  equation (\ref{eqn:entropic_attach}). The validity of this expansion is not valid for $k \ll 2mt$ as the series for $log(x)$ converges very slowly as $x \rightarrow 0$. However at early times after the introduction of the node into the graph, $\frac{k}{2mt}$ will be closer to $1$ and we can expand the $log$ to yield:

\begin{equation*}
    \log  \bigg ( \frac{k}{2mt} \bigg ) \approx \frac{k}{2mt} -1 + \mbox{higher order terms.} 
\end{equation*}

For the same period of time this expression is valid we can see that the leading terms in this expansion contribute to the ODE time evolution of $k$ the following:

\begin{equation}\label{eqn:approx_dk}
    \frac{ d k }{ d t} \approx \frac{\epsilon k}{2t} - \frac{\epsilon k^{2} }{2mt} + \mbox{ higher order terms.}
\end{equation}

What can be asserted is that for a period of time after a node is introduced into the network its behavior will be governed by the first terms in this expansion, with much more complex behavior as the network evolves. This is illustrated nicely in Figure \ref{fig:comp} obtained from our numerical simulations. These first two terms in the expansion are identical in form to the evolution of $k$ with time in the Barab{\'a}si-Albert model, \emph{and also} a correction identical in form to our constrained model. This would indicate that for small $t$ the behavior of the entropic model should closely resemble scale free, with a correction for constraints. As $t$ increases the model will become more complex.

The model introduces $\epsilon$ as a free parameter, and it is a legitimate question to ask what the correct value of this should be. In the numerical simulations we chose, for illustrative purposes, $\epsilon=0.1$. The choice of $\epsilon$ will have a profound affect on the family of graphs that can emerge from the initial conditions and in particular the slope of the power law degree distribution obtained. For example, values of $\epsilon>1$ will tend to generate power laws with $\gamma<3$, and conversely $\epsilon < 1$ will produce $\gamma > 3$, at least in the regime where the first term of equation(\ref{eqn:approx_dk}) dominates. Given that the origin of the parameter is in the relative entropic force of the graph compared to a randomly picked node of degree $k$, one could speculate that its value measures the relative affect of an additional link on the bulk of the graph to increase entropy compared to an individual node of varying degree. High values of $\epsilon$ perhaps indicate relatively more homogeneous graphs than low values, indicating that degree distributions drop off more slowly the more ordered a graph's initial state. In future work we intend to investigate the dependency of graph evolution on $\epsilon$ in more detail, and whether the more complex evolution behavior of our dynamic model has utility in revealing more detail on the internal structure of dynamically evolving graphs.

We believe that there is a deep connection between vertex entropy and the evolution of networks. 
An attractive feature of our model is that it predicts scale free and more complex network evolution behavior from a first principles argument without appeal to any heuristics, node by node parameters, or indeed a stated but not justified property of nodes to seek out other high degree nodes with which to preferentially attach. Instead we argue from the safety of the second law of thermodynamics to a model which reproduces the essential features of scale freedom, and also the constrained  model which we demonstrated provides a better fit to the experimental data.
It is possible that higher terms in the expansion of equation (\ref{eqn:entropic_attach}) could yield insight into the detailed evolution of networks, and provide powerful analytical tools to for example determine the age of a network. Nevertheless, it is attractive to speculate that scale freedom, and similar models, may be a manifestation of the second law of thermodynamics as applied to graph evolution.

Beyond investigating the entropic model, there are many potential enhancements to the constrained model. In further work we intend to conduct analysis of more network datasets and also investigate corrections to the constrained model to improve our estimate of $(c-2m)$ or $(c-\langle k \rangle)$ for the average occupancy of a node, by iterating the resultant distribution in equation (\ref{eqn:main_result}) to calculate $\langle k \rangle$ as $\langle k \rangle = \int_{-\infty}^{+\infty} k P(k) dk$.

\section*{Author contribution statement}
P.T., I.J.W. and G.P. developed the initial model and wrote the initial draft of the paper. I.Z.K. and P.T. refined the exposition of the constrained attachment and continuum models and resolved some technical issues. J.H.P.D. provided subsequent analysis and the results shown in Figure 6. All authors contributed to editing and production of the final manuscript.

\appendix
\section{ - Derivation of $\gamma$ in Constrained Attachment}
\label{app:gamma_deriv}

We recall from the main body of Section \ref{sec:constraints} our expression for $P(k)$ in Equation (\ref{eqn:main_result}):

\begin{equation*}
    P(k) = \frac{2 (c+\delta)\rho^{2/\alpha}t}{\alpha(t+m_{0})} \Bigg(  \frac{ (c+\delta-k)^{\frac{2}{\alpha}-1}} {k^{ \frac{2}{\alpha}+1}} \Bigg) \sim \frac{1}{k^{\gamma}} \mbox{ .}
\end{equation*}

We can simplify this by collapsing the uninteresting details as follows:

\begin{equation}
\label{eqn:simplified_pk}
\begin{split}
    P(k)=\frac{ A(B-k)^{\frac{2}{\alpha} - 1} }{ k^{\frac{2}{\alpha} + 1} } \mbox{, where } \\
                \\
                A=\frac{2 (c+\delta)\rho^{2/\alpha}t}{\alpha(t+m_{0})} \mbox{, and } B=(c+\delta)
\end{split}
\end{equation}

Now, as $a^b=\exp\{ b \log(a) \}$, we can write $(B-k)^{\frac{2}{\alpha} - 1} = \exp \{ (\frac{2}{\alpha} - 1) \log (B-k) \}$. Substituting back into Equation (\ref{eqn:simplified_pk}), and taking the logarithm of both sides, we obtain:

\begin{equation*}
    \log( P(k) ) = \log A + \Big (\frac{2}{\alpha} - 1 \Big ) \log (B-k) - \Big(\frac{2}{\alpha}+1 \Big) \log( k ) \mbox{ . }
\end{equation*}

We can further simplify by noting that $\log (B-k) = \log \{ B( 1-\frac{k}{B}) \} = \log B + \log \Big( 1-\frac{k}{B} \Big)$. We note that if $k \ll B$, either by taking small values of $k$ or allowing $c \rightarrow \infty$, then 
$\frac{k}{B} \rightarrow 0$, so that $\log (B-k) = \log B + \log (1 + 0) = \log B$. Bringing this altogether we have:

\begin{equation*}
    \log( P(k) ) = \log A + \Big (\frac{2}{\alpha} - 1 \Big ) \log B - \Big(\frac{2}{\alpha}+1 \Big) \log( k ) \mbox{ . }
\end{equation*}

Taking the exponential of both sides we end with the main result:

\begin{equation}
\label{eqn:pk_prop}
\begin{split}
    P(k)=\frac{2 (c+\delta)\rho^{2/\alpha}t}{\alpha(t+m_{0})} \times \frac{(c+\delta)^{(\frac{2}{\alpha} - 1)} }{ k^{(\frac{2}{\alpha} + 1)} } \mbox{ , }\\
    \mbox{which is of the form, }\\
    P(k) \propto \frac{1}{ k^{(\frac{2}{\alpha} + 1)} } \mbox{ . } 
\end{split}
\end{equation}

In Equation (\ref{eqn:pk_prop}), we arrive at the familar form of a scale free distribution with $\gamma=2/\alpha+1$. It is interesting to note that, as $c > 2m$, by definition, $\alpha \geq 1$ with equality in the limit that $c \rightarrow \infty$. This yields a range for the power law exponent $\gamma$ as $1 \leq \gamma \leq 3$, with the familiar result of $\gamma=3$ recovered in the case of the constraint being infinite, and therefore unimportant to the dynamics of the network growth.

We can also examine Equation (\ref{eqn:simplified_pk}) in the asymptotic limit of $c \rightarrow \infty$. We recall that $\rho=\frac{m}{c+\delta-m}$, and that $\alpha=\frac{c+\delta}{c-2m}$. At the limit $c \rightarrow \infty$, $\alpha=1$, which reduces Equation (\ref{eqn:simplified_pk}) to:

\begin{equation}
\begin{split}
    P(k) \approx \frac{2c(\frac{m}{c})^{2}t}{(t+m_{0})} \times \Bigg\{ \frac{c}{ k^{3}} - \frac{1}{k^2} \Bigg \} \mbox{ , }\\
    \mbox{which multiplying out and allowing $c \rightarrow \infty$, gives}\\
    P(k) \approx \frac{2m^{2}t}{(t+m_{0})} \times \frac{1}{ k^{3}} \mbox{ .}
\end{split}
\end{equation}

As expected, this is precisely the form of the degree distribution in the standard preferential attachment model, which emerges as the constraint becomes infinite, and therefore unimportant in the dynamical growth of the network.

\bibliography{EPJBConstraints}
\end{document}